\PassOptionsToPackage{table,xcdraw}{xcolor}
\documentclass[twocolumn]{aastex631}

\usepackage[table,xcdraw]{xcolor}
\usepackage{changepage}

\usepackage{ulem}

\makeatletter
\newcommand*{\@rowstyle}{}
\newcommand*{\rowstyle}[1]{\gdef\@rowstyle{#1} \@rowstyle\ignorespaces}
\newcolumntype{=}{>{\gdef\@rowstyle{}}}
\newcolumntype{+}{>{\@rowstyle}}
\makeatother

\usepackage{amstext}
\usepackage{amsmath}
\usepackage{amssymb}
\usepackage{apjfonts}
\usepackage{graphicx}
\graphicspath{{./}{figures/}}

\usepackage{savesym}
\savesymbol{tablenum}
\usepackage{siunitx}
\restoresymbol{SIX}{tablenum}

\usepackage{float}
\usepackage{placeins}
\usepackage{hyperref}

\begin{document}
\received{November 3, 2021}
\revised{August 15, 2022}
\accepted{August 22, 2022}
\submitjournal{ApJL}

\shorttitle{Turbulent Metal Mixing}
\shortauthors{Kolborg et al.}

\title{Supernova-driven turbulent metal mixing in high redshift galactic disks: metallicity fluctuations in the interstellar medium and its imprints on metal poor stars in the Milky Way}

\correspondingauthor{Anne Noer Kolborg}
\email{anne.kolborg@nbi.ku.dk}

\author[0000-0001-7364-4964]{Anne Noer Kolborg}
\affiliation{Niels Bohr Institute, University of Copenhagen,Blegdamsvej 17, DK-2100 Copenhagen, Denmark}
\affiliation{Department of Astronomy \& Astrophysics, University of California, Santa Cruz, CA 95064, USA}

\author[0000-0001-9497-1374]{Davide Martizzi}
\affiliation{Niels Bohr Institute, University of Copenhagen,Blegdamsvej 17, DK-2100 Copenhagen, Denmark}
\affiliation{Department of Astronomy \& Astrophysics, University of California, Santa Cruz, CA 95064, USA}

\author[0000-0003-2558-3102]{Enrico Ramirez-Ruiz}
\affiliation{Department of Astronomy \& Astrophysics, University of California, Santa Cruz, CA 95064, USA}
\affiliation{Niels Bohr Institute, University of Copenhagen,Blegdamsvej 17, DK-2100 Copenhagen, Denmark}

\author[0000-0003-0841-5182]{Hugo Pfister}
\affiliation{Department of Physics, The University of Hong Kong, Pokfulam Road, Hong Kong, China}
\affiliation{Niels Bohr Institute, University of Copenhagen,Blegdamsvej 17, DK-2100 Copenhagen, Denmark}

\author[0000-0002-5095-4000]{Charli Sakari}
\affiliation{Department of Physics \& Astronomy, San Francisco State University, 1600 Holloway Avenue, San Francisco, CA 94132 USA}

\author[0000-0003-2229-011X]{Risa H. Wechsler}
\affiliation{Kavli Institute for Particle Astrophysics and Cosmology and Department of Physics, Stanford University, 382 Via Pueblo Mall, Stanford, CA 94305, USA}
\affiliation{Kavli Institute for Particle Astrophysics and Cosmology, SLAC National Accelerator Laboratory, 2575 Sand Hill Road, Menlo Park, CA 94025, USA}

\author[0000-0001-7493-7419]{Melinda Soares-Furtado}
\altaffiliation{NASA Hubble Postdoctoral Fellow}
\affiliation{Department of Astronomy, University of Wisconsin-Madison, 475 N.~Charter Street, Madison, WI 53703, USA}

\begin{abstract} 
The extent to which turbulence mixes gas in the face of recurrent infusions of fresh metals by supernovae (SN) could help provide important constraints on the local star formation conditions. This includes predictions of the metallicity dispersion amongst metal poor stars, which suggests that the interstellar medium was not very well mixed at these early times. The purpose of this {\it Letter} is to help isolate, via a series of numerical experiments, some of the key processes that regulate turbulent mixing of SN elements in galactic disks. We study the gas interactions in small simulated patches of a galaxy disk with the goal of resolving the small-scale mixing effects of metals at pc scales, which enables us to measure the turbulent diffusion coefficient in various galaxy environments. By investigating the statistics of variations of $\alpha$ elements in these simulations, we are able to derive constraints not only on the allowed range of intrinsic yield variations in SN explosions but also on the star formation history of the Milky Way. We argue that the observed dispersion of [Mg/Fe] in metal poor halo stars is compatible with the star-forming conditions expected in dwarf satellites or in an early low star-forming Milky Way progenitor. In particular, metal variations in stars that have not been phase-mixed can be used to infer the star-forming conditions of disrupted dwarf satellites.
\end{abstract}


\section{introduction}
\label{sec:intro}
The distribution of chemical abundances in stars serves as a “fossil record” of the Galaxy's evolutionary history and provides clues about the types of nucleosynthetic processes that occurred early in the history of the Milky Way (MW). Of particular interest for core-collapse supernova (cc-SN) enrichment is the $\alpha$ element composition of Galactic halo stars. In the metallicity regime [Fe/H] $\approx -4.0$ to $-2.0$ \citep[e..g][]{AudouzeSilk1995,Macias2018}, $\alpha$ elements have been found to be pure core-collapse products with a fairly characteristic [$\alpha$/Fe] abundance ratio but with a star to star bulk scatter in their [$\alpha$/H] concentrations \citep[][]{Cayrel2004, Ryan1996, Frebel2015}. The presence of these elements demonstrates that their synthesis operated in a fairly robust manner in the early MW while their abundance dispersions should yield clues about the formation rate of the massive stellar progenitor systems that lived and dwindled before them. The reason is that metal injection rates can help establish how chemically unmixed and inhomogeneous the interstellar medium (ISM) was when these stars formed \citep[][]{AudouzeSilk1995, Krumholz2018}. At later times, these localized inhomogeneities are expected to be smoothed out as more events take place and cc-SN products are given more time to migrate and mix throughout the progenitor galaxy \citep[][]{Avillez2002,Pan2013,Krumholz2018}. 

To this end, a comprehensive understanding of the inhomogeneous enrichment of the ISM holds great promise for our attempts to decipher the MW’s stellar assembly history \citep[e.g.,][]{Wang2021} and, in particular, the origin of halo stars.
The relative contribution between stars formed in-situ \citep[e.g.,][]{Eggen1962} and those formed ex-situ \citep[e.g.,][]{Searle1978} to the MW halo has been long debated. While recent studies of the respective importance of these different star formation (SF) channels to the halo have primarily supported ex-situ assembly \citep[e.g.,][]{Robertson2005, Johnston2008}, a comprehensive picture remains elusive and a series of issues are still pressing. In particular, issues concerning the specific merger and star formation histories (SFH)s of the individual satellite galaxies.

Comprehensive numerical modeling aimed at addressing the mixing of metals in the MW requires a suite of codes to treat the diverse range of physics and timescales at play. Simulations performed in a cosmological context \citep[e.g.,][]{Few2014, Shen2017, Naiman2018} are required to capture the large-scale mixing mechanisms including satellite mergers, gas inflows, galactic winds and fountains, and shearing disk instabilities, all of which are unquestionably important to comprehend the enrichment and dissemination of metals. On the other hand, since cosmological simulations today are limited by resolutions of a few tens to hundreds of pc, they inevitably entail “sub-grid” models for turbulent mixing, SF, and stellar feedback. As such, ample uncertainties still exist at the 100\,pc scale. For turbulent mixing, the trade-offs are even more pronounced and require a large range of scales to be resolved simultaneously \citep[e.g.,][]{Colbrook2017,Avillez2002, Yang2012}. As a consequence, a single simulation of the full problem incorporating all the aforementioned effects would not only be prohibitively expensive, but also difficult to interpret because of the complexity of the interplay between the various physical mechanisms at different scales. Instead, in this {\it Letter} we study the inhomogeneous enrichment of the ISM via a series of numerical experiments that isolate the key processes that regulate the mixing of SN elements into the ISM.

In addition to being computationally feasible, this approach enables a thorough understanding of the relevant processes. To this end, we present models that study the gas interactions in small patches of a galaxy disk in order to study the small-scale mixing effects of metals at pc scales from SN into the ISM. Resolving the cooling length of individual SN remnants allows us to self-consistently capture the mixing of metals in the ISM, which enables us to calculate the turbulent
diffusion coefficient in various galaxy environments. Additionally, investigating the distribution of $\alpha$ elements in these simulations and comparing them with the observational data can, in turn, provide useful constraints on the star-forming conditions \citep{AudouzeSilk1995, Armillotta2018} within the early MW and/or the infalling satellites that built up the stellar halo. Given the relatively nascent stage of modeling in the field, the results we present in this {\it Letter} amount to a considerable improvement in our understanding of the physical mechanism leading to the inhomogeneous enrichment of the ISM and the mixing of $\alpha$ elements in the early MW.

\begin{table*}[t]
\begin{tabular*}{\textwidth}{=c | +c |+c |+c |+c |+c |+c |+c |+c | +c | +c}
Galaxy type & $\Sigma_\text{gas}$ & $\dot{\Sigma}_\text{SFR}$ & $\Gamma$ & $f_{\rm gas}$ & $\rho_0$ & $L$ & $z_{\rm eff}$ & $\langle \sigma_{v} \rangle$ & $t_{r}$ & $\kappa$ \\
 & [$\SI{}{M_\odot \per pc^2}$] & [\SI{}{M_\odot \per kpc^2 \per Myr}]& [\SI{}{/ kpc^2 / Myr}] & & [\SI{}{g \per cm^3}] & [\SI{}{pc}] & [\SI{}{pc}] & [\SI{}{km \per s}] & [\SI{}{\mega yr}] & [\SI{}{pc \kilo \meter \per s}] \\ \hline 
MW progenitor, high SFR & 50 & \num{3e4} & \num{300} & 0.100 & \num{3.47e-23} &$1000$ & 40 & 11 & 19 & 147 \\
MW progenitor, low SFR & 5 & \num{1e3} & \num{10}& 0.088 & \num{2.08e-24} & $1000$ & 80 & 8 & 39 & 213 \\
Satellite galaxy & 5 & \num{8e2} & \num{8}& 0.496 & \num{4.67e-25} & $4000$ & 365 & 12 & 120 & 1460 \\
\rowstyle{\color{gray}}
MW progenitor, low $\epsilon_*$ & 5 & \num{2e2} & \num{2}& 0.088 & \num{2.08e-24} & $1000$ & 80 & 4 & 73 & 107 \\
\rowstyle{\color{gray}}
High redshift, high $\epsilon_*$ & 5 & \num{5.5e3} & \num{55}& 0.088 & \num{2.08e-24} & $1000$ & 80 & 14 & 23 & 373
\end{tabular*}
\caption{Parameters for all the galaxy models examined in this project, columns are: simulation name, gas surface density, star formation rate surface density, surface density of rate of SNe ($\Gamma = \dot{\Sigma}_\text{SFR}/100 M_\odot$), fraction of gas mass to stellar mass, initial density of gas in the mid plane of the disk, simulation box size, effective scale height of the disk, mean velocity dispersion of the gas, characteristic relaxation time and turbulent diffusion coefficient. The results of the supporting simulations (in grey) are presented in Appendix \ref{app:MW_eff}.}
\label{tab:simulations}
\end{table*}

\section{Methods and Metal Mixing Experiments}
\label{sec:sims}
We use the hydrodynamical, adaptive mesh refinement (AMR) code RAMSES \citep{Teyssier2002} to model the gas interactions in small patches of a galaxy disk to study the small-scale mixing effects of metals from SNe into the ISM. The simulation set-up is described in detail in \citet{Martizzi2016} and here we summarize the salient features and provide details for those aspects of the set-up that are different to their investigations. 

The set-up studies small patches of galaxy disks at relatively high resolution. In particular, we are able to capture the momentum injection and thermal energy of individual SNe, as the cooling phase of individual remnants is well resolved. 
By changing the parameters of the gravitational potential, gas fraction, gas density, and star formation efficiency ($\epsilon_*$), we are also able to replicate the characteristics of different local galactic environments that might be representative of the early MW or its accreted satellites. 

For the purposes of this {\it Letter} we have chosen to focus on two different possible MW progenitors as well as one likely satellite candidate. The MW progenitors are chosen to model two different high redshift cases. One is a gas-rich, highly star forming model \citep[][]{vanDokkum2013, Shen2015, Naiman2018}. 
The other one has a lower star formation rate (SFR) and is motivated by the recent findings of \cite{Wang2021}, whereby the presence of the Large Magellanic Cloud (LMC) satellite hints at a MW progenitor with a less vigorous SFR. 
Finally, we include a dwarf-like satellite galaxy in order to examine the viability of an ex-situ birth environment for the halo stars. In order to study how mixing is altered by changes in the gas-to-light ratio, we have also run two additional versions of the low SFR MW progenitor model with the same gas density but with two different star formation efficiencies. The results of these additional simulations are included in Appendix~\ref{app:MW_eff}. Table~\ref{tab:simulations} summarizes the parameters of all the models used in this work. 

The size of the computational domain changes between different models. It increases for the satellite galaxy simulation to accommodate the larger scale height of the disk, which is caused by the reduced gravity of the galaxy. We use a regular Cartesian grid (all cells within the box have the same physical size, but not the same mass resolution) and fix the resolution at $2^8$ cells. This is sufficient to ensure that the cooling radius is resolved by at least a few resolution elements for the majority of the supernova remnants (SNR) in all models. The cooling radius is calculated using \citep[][]{Martizzi2016}: 
\begin{equation}
R_\text{c} \approx \SI{14}{}
\left( \frac{n_\text{H}}{\SI{1}{\centi \meter^3}} \right)^{-3/7} \left( \frac{E_\text{SNe}}{10^{51}{\rm erg}} \right)^{2/7} \left( \frac{Z}{Z_\odot} \right)^{-1/7}{\rm pc}.
\label{eq:Rcool}
\end{equation}
In each model the percentage of SNe that are not resolved by at least 2 (5) cells at their cooling size are: MW progenitor, high SFR 0.6\% (6.1\%); MW progenitor low, SFR 0\% (1.6\%); and satellite galaxy 0\% (3.8\%). 

The SNe are seeded following the "fixed" distribution scheme used in \citet{Martizzi2016}; in this framework a scale height for explosions is defined and the SNe have equal probability of happening anywhere in the volume contained within this height and zero probability outside. In all the models used here, this height is twice the gaseous scale height of the disk ($z_\text{SNe} = 2 z_\text{eff}$); $z_\text{eff}$ is defined in Section~\ref{sec:steady}. 
Each SN injects a mass of $M_\text{ej} = 6.8\,M_\odot$ and an energy of $E_{\rm SNe} = 10^{51}\,{\rm erg}$. 
This energy is partitioned into thermal energy and momentum injection using the sub-grid model of \citet{Martizzi2016}, which also distributes the mass among the neighboring cells of the explosion. The rate of SNe per unit area, $\Gamma$, is given by the SFR as $\Gamma = \frac{\dot{\Sigma}_\text{SFR}}{\SI{100}{M_\odot}}$.

The evolutionary timescale ($t$) of each simulation is dictated by the SFR and differs between simulations in order to account for the gentler evolution of galaxies with lower surface densities of gas (efficient cooling is impaired) and lower rate of SNe (feedback is reduced). The time between consecutive snapshots ($\delta t$) is also adjusted to ensure that there are not too many SNe happening in each snapshot, i.e., models with higher rates of SNe have lower $\delta t$. This allows us to follow the interaction between the remnant and the ISM more closely. The specific time intervals are: MW progenitor, high SFR: $\delta t=$ \SI{5e4}\,{year}; MW progenitor, low SFR: $\delta t=$\SI{5e4}\,{year}; and satellite: $\delta t=$\SI{1e5}\,{year}. 

\begin{figure}
\includegraphics[width = 0.5\textwidth]{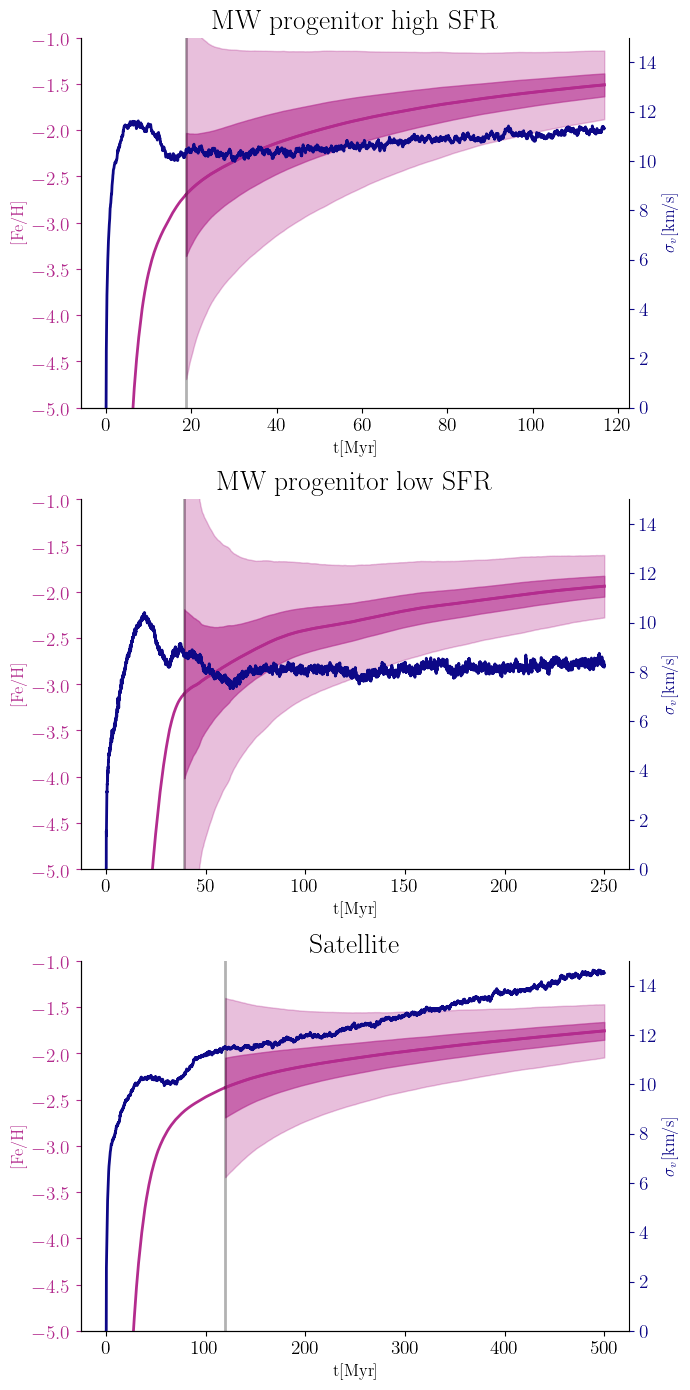}
\caption{Evolution of the velocity dispersion, $\sigma_{v}$ (blue) and the mean [Fe/H] abundance ($\langle [ \mathrm{Fe} / \mathrm{H}] \rangle$) (purple line) as well as $1 \sigma$ and $3 \sigma$ spread (shaded regions) as a function of time. The relaxation time, $t_{r}$, for each model is marked by the vertical gray line in each panel. The panel titles indicate the simulation. Models with lower SFR have a more slowly evolving $\langle [ \mathrm{Fe} / \mathrm{H} ] \rangle$ but tend to have smaller spread at the same mean abundance as models with higher SFR. \label{fig:time_evolution}}
\end{figure}

\subsection{Metal injection} \label{sec:metal_inj}
We monitor individual elements as passive scalars. For the purposes of this {\it Letter} we focus on Mg (as our representative $\alpha$ element, see Appendix \ref{app:obs}) and Fe (as a tracer of the global metallicity), in relation to the background element H. Metals are introduced into the gas by SNe, which are initially assumed to be chemically identical. An assumption that we relax to include intrinsic SN yield variations.

In the first case, each SN yields masses of $M_\text{Mg} = 0.10\,M_\odot$ and $M_\text{Fe} = 0.08\,M_\odot$. These values are chosen to achieve a yield ratio of $\frac{M_\text{Mg}}{M_\text{Fe}} = 1.27$, equivalent to the best fitting value for the observational set we wish to compare with (see Appendix \ref{app:obs}). 
This approach is similar to assuming that the mean abundances are fixed by the mean yields, which is a commonly used conjecture when studying metal mixing \citep[e.g.,][]{Dalcanton2007}. This allows us to isolate the effects that turbulent mixing has on the abundance spread of metals within the ISM. 

After studying the mixing of Fe as a function of the global mean [Fe/H] abundance, $\langle [\mathrm{Fe} / \mathrm{H}] \rangle $, in simulations with no intrinsic spread in SN yields, we then proceed to alter the [Mg/Fe] yields. We do this by introducing an intrinsic spread in yields from SN to SN. This assumption is supported by our finding in Figure~\ref{fig:mass_fit} that the spread in [Mg/Fe] can only be produced by altering the SN yields. We model the yield variation by introducing two types of SNe with different [Mg/Fe] yields. We choose these yields such that their mean corresponds to the mean [Mg/Fe] abundance of the observational data set in the metallicity range $-3.5 < [\mathrm{Fe} / \mathrm{H}] \leq -1.0$. This range of metallicity was selected in order to exclude a handful of very low metallicity stars with large associated uncertainties and to ensure that the chosen range is wider than the range of metallicities within which the simulations are evaluated (Figure~\ref{fig:time_evolution} and Section~\ref{sec:stars}) The higher yield value in the spread is chosen to be similar to the most highly enriched star in the sample, and the lower yield value is then fixed such that the desired mean is achieved. This results in yield values of [Mg/Fe] = 0.72 and [Mg/Fe] = -1.39 for the two types of SN injection sites.

The reader is referred to Section~\ref{sec:stars} for a discussion on the implications of our findings in the context of abundance determinations of metal poor stars. One key aspect of the comparison between our simulations and the observations is that we are directly comparing how the spread in [Mg/Fe] evolves with the $\langle [ \mathrm{Fe} / \mathrm{H}] \rangle$, which is initially sensitive to the assumed SN-to-SN yield variance and is then subsequently smoothed out by turbulent diffusion.\\

\section{Simulating Supernova Feedback and Mixing in galaxy disk patches}

\subsection{Steady State} 
\label{sec:steady}
The early time of the simulations are characterized by a relaxation from the initial conditions. Initially the disk is supported by thermal pressure, while at later times the disk is supported in part by turbulent pressure. As a result, the true effective
scale height differs slightly from what one would calculate assuming hydrostatic equilibrium with the unchanging gravitational potential. To address this, \citet{Martizzi2016} defined the effective scale height, $z_\text{eff}$ as: 

\begin{equation*}
 z_\text{eff} = \frac{\Sigma_\text{gas}}{2 \rho_0} = \frac{1}{2 \rho_0} \int_{-L/2}^{L/2} \rho_{\text{gas}}(z) \, dz,
\end{equation*}
where $\Sigma_\text{gas}$ is the surface density of gas and $\rho_0$ is the gas density in the disk midplane during initial conditions. Both $\rho_0$ and $z_\text{eff}$ are listed in Table~\ref{tab:simulations} for all models. 

We use $z_\text{eff}$ to define a characteristic relaxation time, $t_{r}$, for our disk models:
\begin{equation}
 t_{r} = \frac{4 z_\text{eff}}{\langle \sigma_v \rangle},
 \label{eq:tr}
\end{equation}
$\langle \sigma_v \rangle$ is the time averaged, mass-weighted velocity dispersion of the gas. We calculate the mass weighted velocity dispersion as:
\begin{equation*}
 \sigma_v = \sqrt{\frac{\sum_i^N (\vec{v_i} - \langle \vec{v} \rangle)^2 M_i}{\sum_i^N M_i}},
\end{equation*}
where $N$ is the number of cells in the simulation volume, $M_i$ is the mass of gas in the cell, $\vec{v}_i$ is the gas 3D velocity, and $\langle v \rangle$ is the mean velocity of the gas. The parameter $\langle \sigma_v \rangle$ is set by the properties of the turbulence in the disk and depends weakly on the surface density of gas \citep{Martizzi2016}. The evolution of the velocity dispersion is shown in Figure~\ref{fig:time_evolution}. 
We calculate the mean of the velocity dispersion over the entire evolution time of the model and report this value in Table~\ref{tab:simulations}. 

The factor of $4$ in equation~\ref{eq:tr} is a choice we made to ensure that information has been communicated effectively to all regions and that turbulent motions in the bulk of the disk
reach a statistical steady state \citep{Martizzi2016}. This choice is validated by comparing the evolution of the velocity dispersion before and after $t_{r}$, in almost all cases the velocity dispersion has a fairly gradual (or even flat) development after the relaxation time and a much more volatile development before (Figure~\ref{fig:time_evolution}). This dynamical timescale results from complex interplay between the gas density, the shape of the gravitational potential, and the SFR. For instance, when the gas density increases, the disk tends to relax more promptly, because as gas cools it comes into a quasi-equilibrium with the gravitational potential and higher density gas will cool more efficiently thus reaching steady state sooner. This process is nicely exemplified in the top and middle panels of Figure \ref{fig:time_evolution}. In Appendix \ref{app:MW_eff} we compare models with the same gas density and gravitational potential but varying SFR, here we find that $t_r$ increases with increasing SNe feedback. In all cases we find that steady state is reached at $t \approx t_{r}$. In the remainder of this work, we only show results for $t > t_{r}$, when turbulent motions in the bulk of the disk have reached a statistical steady state.

In both MW progenitor cases, the velocity dispersion flattens out at or soon after the relaxation time, indicating that these models are dynamically stable over long periods of time. The dwarf model however shows a continually increasing velocity dispersion with time, indicating that this model is in fact not dynamically stable. 
Over the course of the simulation, the dwarf model looses $\approx 35\%$ of its initial mass in SNe driven winds. In comparison, the two MW progenitor models lose $\lesssim 5\%$. This behaviour is consistent with what is routinely observed when modelling dwarf galaxy systems \citep[e.g.,][]{Fragile2003,MacLow1999, Fielding2017}. We see similar behaviour in the MW progenitor with low SFR and high $\epsilon_*$ (see Appendix \ref{app:MW_eff}), which loses $\approx 10\%$ of its mass to SNe driven winds during its lifetime. 

\subsection{Connecting metal abundance variations to star-forming gas conditions}
\label{sec:metal_evo}

\begin{figure*}
 \centering
 \includegraphics[width = \textwidth]{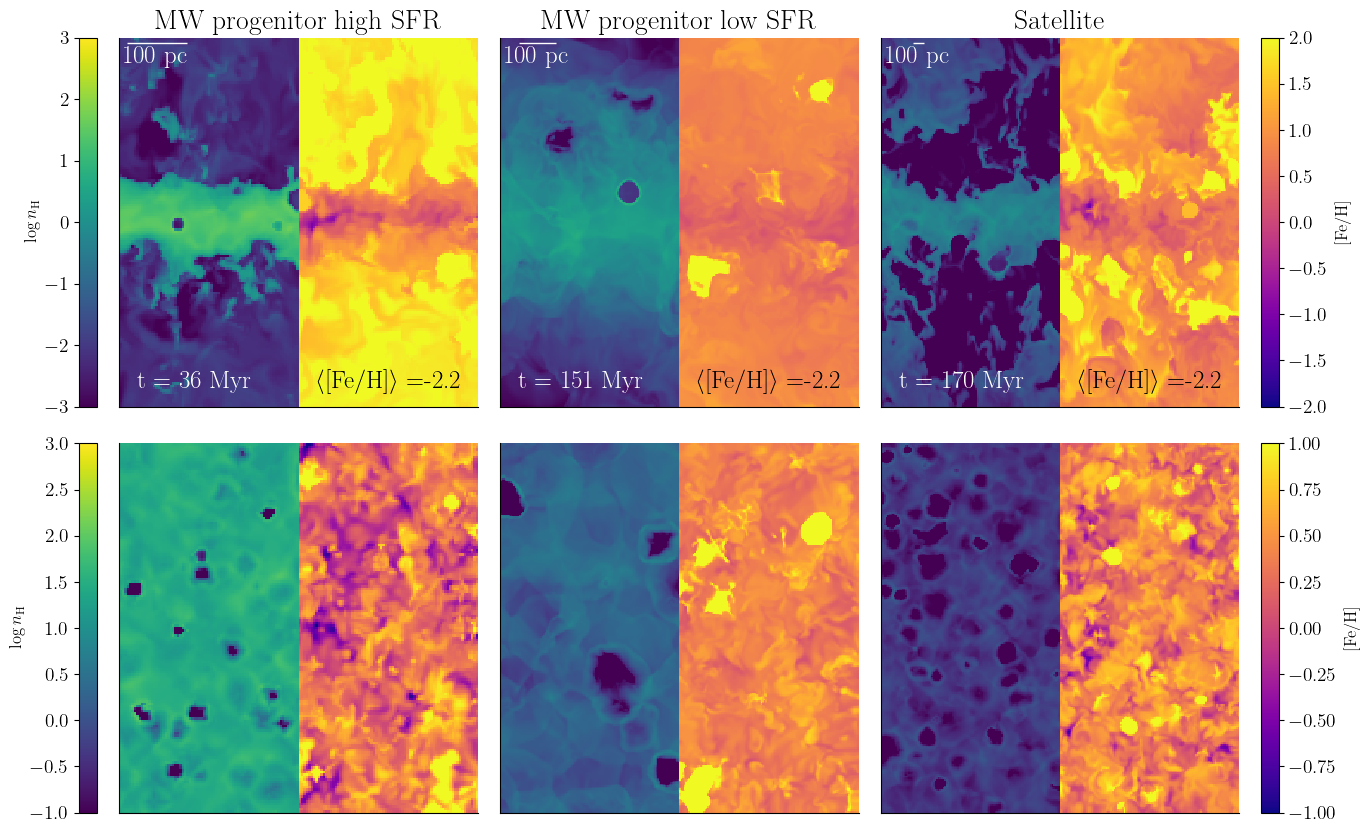}
 \caption{Slice maps through the center of the computational domain for different galaxy models, indicated by the panel titles. All snapshots are taken when $\langle [\text{Fe}/\text{H}] \rangle$ = -2.2. The number of SNe, $N_\text{SNe}$, required to reach this mean abundance is ~5600, ~1300, and ~14,000, respectively. The top row shows an edge on view, while the bottom row shows a face on view of the disk. In each panel, the left-hand side (blue-green color map) is the number density of H ($\log_{10} n_\text{H}$) while the right-hand side (purple-yellow) is the local gas abundance of [Fe/H]. Each panel is $6 z_\text{eff}$ along each axis, while the scale bars indicate the relative sizes between panels. The time since the beginning of the simulation is shown. These times correspond to $1.89 t_{ r}$, $3.87 t_{r}$, and $1.41 t_{r}$, respectively. In all models, individual SNR are clearly resolved and the increased homogeneity of the lower SFR models at the same $\langle [\text{Fe}/\text{H}] \rangle$ can be easily observed.}
 \label{fig:maps}
\end{figure*}

Over time, the material within the disk is progressively enriched as subsequent SNe increase the mean metallicity of the gas. At the same time, SNe drive turbulent mixing, which causes the metals to spread from their localized injections sites. In general, models with fewer SNe take longer to arrive at the same mean metallicity.
As such, their products are given more time to migrate throughout the disk. This implies that galactic disks with lower SFRs show more homogeneity at a comparable average abundance. This can be seen in the evolution of $\langle [\text{Fe}/\text{H}] \rangle$ shown in Figure~\ref{fig:time_evolution}.

[Fe/H] is often used, for convenience, as a rough proxy for age when thinking about the assembly of metal poor stars \citep[e.g.,][]{Cayrel2004,Frebel2010}. This relation is shown in Figure~\ref{fig:time_evolution} and is, as expected, non-linear. 
Thus, a given galaxy disk patch spends significantly longer time at higher abundance values than at lower ones. This relationship is useful to keep in mind as we study the scatter in [Fe/H] of the gas at a given $\langle [ \mathrm{Fe} / \mathrm{H} ] \rangle $.

In all simulations, an early, chemically unmixed and inhomogeneous galactic disk is observed. As time evolves, these localized inhomogeneities within the gas are smoothed out as more events take place and SN products are granted more extended periods to migrate and mix across the disk. Figure~\ref{fig:maps} shows slice maps for each of the three galaxy models considered in this work. Individual SNR are clearly visible as pockets of lower density (higher pressure) gas with high metal abundances. The influence of the density and density gradients is also clearly seen when comparing between different models. The lower density gas of the MW progenitor with low SFR allows each remnant to reach larger physical sizes than remnants in the higher density environment of the MW progenitor with high SFR, leading to a higher degree of homogeneity in the [Fe/H] abundance of the gas at given $\langle [\text{Fe}/\text{H}] \rangle$ in the lower SFR models. In Appendix~\ref{app:MW_eff} we show that for identical surface density of gas (and galaxy potential) a lower rate of SNe also leads to a higher degree of homogenization of the gas at a given mean metallicity, which is caused by the slower evolution with time, giving each SNR longer to evolve before the required metallicity is reached. 

In essence, we see a direct connection between the rate of metal enriching events (which follows the SFR) and the degree of spread in the abundance of the ISM at a given global average metallicity. As the SFR increases, so does the extent of metal disparity across the simulation. This is because newly formed gas is injected at a faster rate than the rate at which turbulent mixing smooths them down.

The nearly uniform density along the disk plane compels the remnants to remain nearly circular, while the density gradient perpendicular to the disk causes them to expand more easily along the steep vertical pressure gradient. This allows metal enriched outflows to reach the edge of the disk and be ultimately expelled \citep{Martizzi2015}. A review of the time series of these metal maps shows enriched gas getting expelled from the disk, rising within the gravitational potential before cooling and falling back, which is reminiscent of a mini galactic fountain. We note here that some gas is able to leave the simulation domain, and the outflow boundary conditions at the top and bottom of the computational box does not allow it to return. As a result, enriched material is steadily lost from the galactic disk patch. As discussed in Section \ref{sec:steady} this is a small effect for the MW progenitor models but a considerable one for the dwarf galaxy model which cannot be considered dynamically stable over longer periods of time and is unable to effectively retain the metals injected by SNe. In Section \ref{sec:steady} and Appendix \ref{app:MW_eff} we show that the MW progenitor model with low SFR and high $\epsilon_*$ exhibits the same behaviour, although to a lesser degree than the dwarf-satellite model. 

Having discussed the evolution and the salient features of the small patch disk simulations, we now turn our attention to quantifying the degree of metal mixing and its relevance to spectroscopic observations of stars in the MW.

\section{Metal Mixing and Comparison to Observations of Metal Poor Stars}
\label{sec:mixing}
In this section we examine the mixing of metals by SNe in the surrounding gas. In Section \ref{sec:loc_mix}, we consider the local mixing around the sites where SNe, assumed to be chemically identical, recently occurred. In Section \ref{sec:stars}, we study the spread in [Mg/Fe] abundances achieved in each of our models with SN metal injections with intrinsic variations in yields. In this case, we compare the results of the simulations with the [Mg/Fe] spreads inferred from metal poor halo stars in the MW with the goal to constrain the SFR in the birth environments of these stars.

\subsection{Understanding local mixing} 
\label{sec:loc_mix}
In this section we leverage the high spatial and temporal resolution of the patch disk simulations to study the metal mixing properties of individual SNe. In each simulation, we record the time and position of all SNe, thus enabling us to study their evolution and the associated metal dispersion across the galaxy disk.

\subsubsection{Metal mixing at the cooling scale}
\label{sec:cool_mass}

\begin{figure}
 \centering
 \includegraphics[width = 0.5\textwidth]{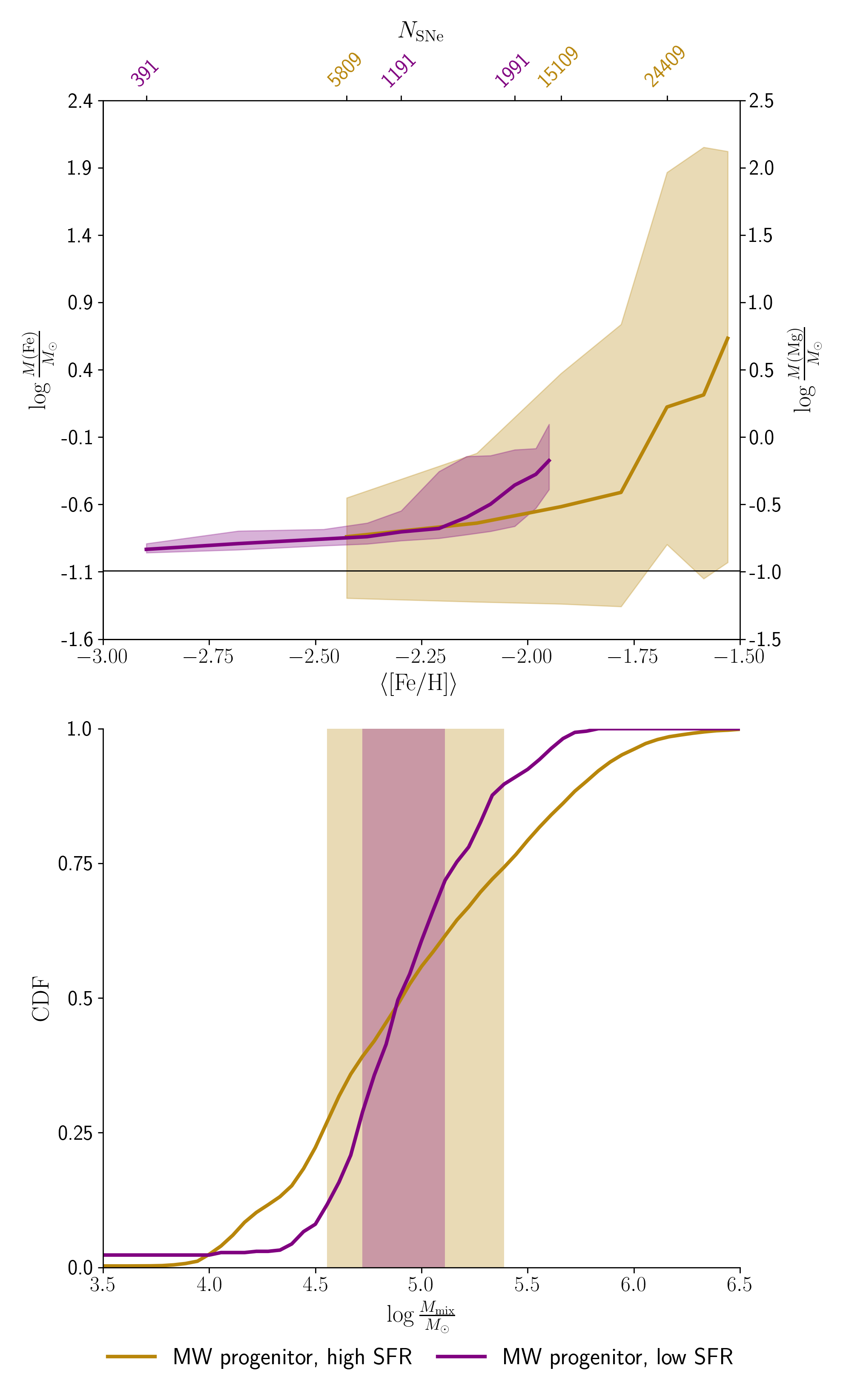}
 \caption{Top panel: Median (solid lines), 25th, and 75th percentile (shaded region) of the mass of Fe (Mg) enclosed within $R_\text{cool}$ around all SN injection sites (average over \SI{15}{Myr} intervals) in the MW progenitor with low SFR (purple) and the one with high SFR (yellow). The solid black line marks the mass of Fe per injected SN and thus provides a measurement of the degree of overlap between adjacent mass injection events. Both models are marginally consistent with a single enrichment model at low $\langle [ \mathrm{Fe} / \mathrm{H}] \rangle $. Bottom panel: cumulative distribution functions (CDF) of the mixing masses derived for all injected SNe in the two MW progenitor models (colors are the same as top panel). Shaded region denotes the IQR for both distributions. The patchy distribution of elements in the higher SFR model translates to a wider distribution of mixing masses for the SNe.
 \label{fig:mixing}}
\end{figure}

The origin of the dispersion of Fe (Mg) masses around injection sites depends sensitively on the degree of overlap between the metals injected from neighboring SNR. Intuitively, a population of chemically identical SN is well mixed at a particular length scale when the dispersion of Fe (Mg) masses measured across the disk at this particular scale is modest. Here, we select the cooling radius around all of the injection sites as our characteristic scale. In order to integrate the metals into new stars, the swept up mass must first cool \citep{Nomoto2013,Macias2018}; therefore, the cooling mass yields a robust lower limit on the minimum amount of enriched gas that can be incorporated into the next generation of stars. The metal content within a cooling mass thus provides a simple test of the single enrichment star formation assumption \citep[e.g.,][]{Karlsson2013}.
It is useful to recall that the cooling radius (and hence the cooling mass) depends sensitively on the mean density of the surrounding gas (Equation~\ref{eq:Rcool}) but only weakly on its mean metallicity \citep{Karpov2020}. 

The upper panel of Figure~\ref{fig:mixing} shows the enclosed Fe (Mg) mass within the cooling scales of the SNe in both MW progenitor models. Both models display an initial quasi-linear trend in the Fe (Mg) mass with metallicity, followed by a change of slope at $\langle [\text{Fe}/\text{H}] \rangle \approx -2$. The early behavior follows the expected evolution for overlapping SNR expanding in a medium in which the enrichment of the surrounding material is dominated by recently injected metals and not by the mass of the metals residing in the ISM. At $\langle [\text{Fe}/\text{H}] \rangle \approx -2$, the mass of Fe (Mg) contained within the swept-up ISM begins to be a meaningful contribution to the total Fe (Mg) mass within a cooling radius. 

In this formalism, the number of overlapping SN within a cooling mass can be easily inferred by comparing the total enclosed mass of any element with that expected from a single SNR, which is illustrated (for Fe) by the solid black line (upper panel of Figure~\ref{fig:mixing}).

As expected, the dispersion of (Mg) within a cooling mass is smaller for the low SFR model. When examining the morphology of the individual SNR in Figure \ref{fig:maps} it is clear that this happens because neighboring remnants in this model show significant overlap. This effect is compounded by the larger cooling radii of the remnants, caused by the lower mean density of the surrounding ISM.
In contrast, the high SFR model, despite having a larger number of injected SNe at the same $\langle [\text{Fe}/\text{H}] \rangle$, shows a larger dispersion in the number of SNR that overlap at the chosen scale. Across the disk, we see regions with no overlapping remnants, while other regions show significant overlap between neighboring remnants. This emanates by virtue of the fact that individual remnants in this model remain compact enough to be restricted to scales smaller than the disk scale height. As such, individual remnants sample a wide range of density scales and thus a corresponding wide range of cooling masses. 

\citet{Macias2018} argued that the abundances of [Mg/H] of stars with $[ \mathrm{Fe} / \mathrm{H} ] \lesssim -2$ could be consistent with enrichment from SN sites yielding $M_\text{Mg} = 0.1\,M_\odot$. Our results suggest that while it is unlikely (at the SFRs considered here) to form stars from gas that has been enriched by a single event, stars with enrichment by a few events are plausible at $\langle [\mathrm{Fe}/\mathrm{H}] \lesssim -2$. This is in broad agreement with the results of \cite{Welsh2021} who modelled the stochastic enrichment of metal poor stars by the first SNe and found the expected number of enriching first SNe to the most metal poor stars to be $N_{\rm SNe} = 5_{-3}^{+13}$. 

Our simulations suggest that the ISM in the galactic disk patches increased rapidly in metallicity after less than 100\,Myr. With the assumption that Pop III stars can only form in gas with metallicity below a critical value, it is thus worth quantifying the pristine gas fraction below a threshold metallicity. Following \citet{Pan2013}, we calculate the fraction of gas with $[ \mathrm{Fe} / \mathrm{H} ] < -6$ at the onset of steady state in each of our models and find $\lesssim 0.5$\% (high SFR MW progenitor at $t_r \approx \SI{20}{Myr}$), $\lesssim 1.5$\% (low SFR MW progenitor at $t_r \approx \SI{40}{Myr}$) and 0\% (dwarf-like satellite at $t_r \approx \SI{120}{Myr}$). The ubiquity of heavy elements in the Lyman alpha forest indicates that there has been widespread diffusion from the sites of these early SNe, which is consistent with our findings. \citet{2003ApJ...586....1M} estimated that most of the star-forming material in the intergalactic medium (IGM) has reached $[ \mathrm{Fe} / \mathrm{H} ] < -3.5$ in the range $z\approx 15-20$. Constraining how these Pop III stars form at high-$z$ is thus of great relevance.

\subsubsection{Characteristic mixing masses}
\label{sec:mix_mass}
Enrichment of the ISM depends on how efficiently metals are mixed with the ambient material swept up by the individual blast waves. At the time of shell formation (i.e., cooling radius), we find the mass fraction of Fe (Mg) to be enhanced in relation to the surrounding medium for most injected remnants, although that fraction decreases as the simulation evolves. These localized Fe (Mg) enhancements are then smoothed out as the enriched gas is given more time to expand and mix throughout the disk.

Motivated by this, we calculate the scale at which the enclosed [Fe/H] ([Mg/H]) around each SN site is less than or equal to the mean abundance,$\langle [\text{Fe}/\text{H}] \rangle$ ($\langle [\text{Mg}/\text{H}] \rangle$), of the disk. We refer to this integrated mass as the mixing mass, which broadly represents the characteristic scale at which SNR become chemically indistinguishable from the background gas. The bottom panel of Figure~\ref{fig:mixing} shows the cumulative distributions of the mixing masses in each of the two MW progenitor models. A modest fraction of all SNR reach their mixing masses within the scale height of the disk, only those that do are shown. The fraction of the remnants that are confined within $\lesssim 2 z_\text{eff}$ of the disk is 25\% for the MW progenitor model with high SFR and 20\% for the model with low SFR.

The MW progenitor with low SFR shows a narrower range of mixing masses, which is indicative of a more homogeneous distribution of metals within the disk. By contrast, the high SFR model exhibits a more patchy distribution of elements at the same $\langle [\text{Fe}/\text{H}] \rangle$, leading to a larger spread in mixing masses. The smallest mixing masses within the distribution can be understood as the masses of those SNe that happened to go off in more pristine environments at a time when the disk has already achieved a large mean metallicity, so less dilution is required to reach the background value. Conversely, the highest mixing masses are caused by SNe that occur in regions where the local abundance is higher than the disk average background abundance and are thus harder to dilute. 

It is noteworthy that the range of mixing masses we derive from these models are similar to the range obtained by \citet{Macias2018} for metal poor stars in the same [Fe/H] range, which were calculated under the assumption of mixing with a perfectly pristine background. This assumes that the cold gas that turns into new stars preserves the metal variations in the gas at a given $\langle [\text{Fe}/\text{H}] \rangle$. 
Having made the case that the spread of stellar abundances at a given $\langle [\text{Fe}/\text{H}] \rangle$ can be associated with the star formation properties of the progenitor disk and their end products, it is worth thinking further about the implications.

\subsection{Starforming gas and comparison to observations}
\label{sec:stars}

 \begin{figure*}
 \centering
\includegraphics[width = \textwidth]{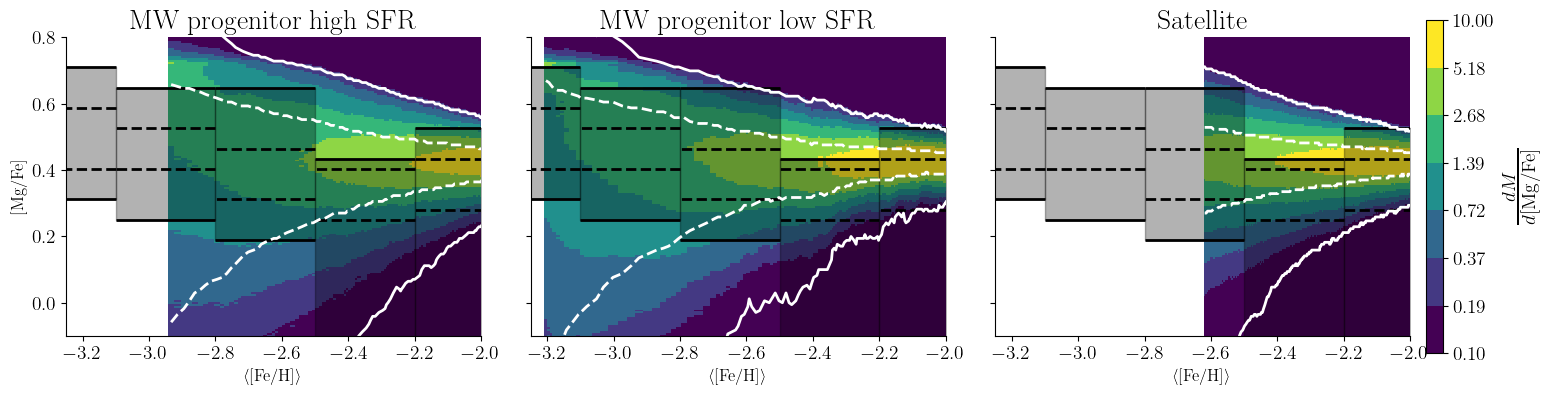}
\caption{Mass weighted distributions of [Mg/Fe] in the cold, dense gas as a function of $\langle [\mathrm{Fe}/\mathrm{H}] \rangle$ for SN injection sites with intrinsic yield variations. The solid white contours indicate the 99\% extent of the distributions while the dashed lines indicate the 68\% extent. The black shaded regions indicate the [Fe/H] bins used to bin the abundances measured by \citet{Roederer2014}, the solid black lines indicate the 99\% span of the sample in the bin and the black, dashed lines indicate the 68\% span. From left to right, the surface density of gas in the modelled galaxy decreases.}
\label{fig:cold_dense_abundances}
\end{figure*}

Motivated by the results presented in the previous section, we examine the spread of [Mg/Fe] as a function of $\langle [\text{Fe}/\text{H}]\rangle$ predicted by our models, under the assumption of intrinsic yield variations. This allows us to directly compare our models to observations, given that we have two independent tracers of metal mixing. One being the spread in [Mg/Fe], which is set at the onset by the assumed SN-to-SN yield variance and is then subsequently
smoothed out by turbulent diffusion, and the other one being the average $\langle [\text{Fe}/\text{H}]\rangle$, whose evolution is driven by metal injection and turbulent diffusion. \\
Our framework does not self-consistently incorporate star formation so, in order to make a meaningful comparison, we consider the cold gas within the disk that is the most likely to form stars. We impose a temperature threshold of $T \leq 10^{4}\,{K}$, associated with the cooling function temperature floor, and a density threshold of $\rho \gtrsim (\rho_0/4)$. At each simulation output we calculate the mass-weighted distributions of [Fe/H] and [Mg/Fe] in the gas that fulfills both of these requirements.\footnote{We note that this leads to a slightly different mean abundance than what is considered in the previous Sections where all the gas, not only the cold component, is used for the calculation.}

In Figure~\ref{fig:cold_dense_abundances}, we show the distributions of [Mg/Fe] as a function of $\langle[ \mathrm{Fe}/ \mathrm{H}] \rangle $. The color map indicates the gas mass ($M_\odot$) in each bin of [Mg/Fe] abundance. The differences in $\langle [\mathrm{Fe} / \mathrm{H} ] \rangle$ range covered in each panel are due to the different speeds at which the patches are enriched by SNe in combination with the different time scales over which the models come to steady state. All models are initialized at the same $\langle [ \mathrm{Fe} / \mathrm{H} ] \rangle$ but as some models relax more quickly the first distribution included may be at lower metallicity compared to a model which relaxes more slowly (see Section \ref{sec:steady}). The solid white contours indicate the 99\% extent of the [Mg/Fe] distributions, while the dashed white lines denote the 68\% extent. The metal poor halo stars used for this comparison are taken from \citet{Roederer2014} and the specific sample selection criteria is discussed in Appendix~\ref{app:obs}. We opt to compare our simulations with this sample because of its relatively large size ($\approx 300$ metal poor halo stars) and the care taken by the authors to document the uncertainty of their measurements. This allows us to use the star-to-star [Mg/Fe] scatter as a function of [Fe/H] derived by the authors as a constraint for the intrinsic [Mg/Fe] elemental variations driven by turbulent diffusion.

Some obvious points should be emphasized. The degree of metal mixing of the cold gas within the disk is, as we argue in the preceding sections, directly impressed on the extent of [Mg/Fe] spread. This spread arises naturally from the intrinsic distribution of Mg and Fe yields in SNe \citep[e.g.,][]{Kobayashi2006} and is clearly seen in the abundance determinations of metal poor stars. In the simulations, this metal spread is initially set by the selected yield variance and is then gradually smoothed down by turbulent diffusion. We note that the exact shape of the simulated distribution of [Mg/Fe] depends sensitively on the assumed SNe yields, while the shape of the outer envelope is primarily determined by turbulent diffusion, whose coefficient is sensitive to the properties of the galaxy (Table \ref{tab:simulations}). As a result, the evolution of the spread of [Mg/Fe] with $\langle[ \mathrm{Fe}/ \mathrm{H}] \rangle $ is
an observable diagnostic of the star formation processes occurring within the disk.

In almost a literal sense, the stage is set for us to interpret the observational properties concerning metal poor stars in the context of turbulent mixing driven by SN feedback. Figure~\ref{fig:cold_dense_abundances} shows that the simulations provide us with information about how the derived [Mg/Fe] spreads in the simulations evolve with $\langle [ \mathrm{Fe} / \mathrm{H}] \rangle$, which depends on the star forming properties of the system. As such, we might be able to constrain the SFR of the birth environments of metal poor stars when the derived [Mg/Fe] spreads in the simulations are unable to effectively describe the observed distributions. 

There are, however, three effects that renders the task of interpreting the simulations in the context of observations challenging. The first stems from the assumed SNe yield distributions, which are not well constrained and are likely sensitive to the IMF and the explosive properties of the progenitors \citep[e.g.,][]{Arnone2005}. A second challenge is that the assembly process that led to the metal poor stars we observe today likely involves multiple populations \citep[e.g.][]{Bullock2001, Robertson2005, Johnston2008, Deason2016}. Third, our understanding of the Milky Way's stellar halo based on the analysis of local stellar samples is yet incomplete due to the limited sampling generated by the survey footprints. In particular, the observational sample used in this analysis suffers from incompleteness and selection effects, which are most prominent at lower [Mg/Fe] abundances \citep[][see also Appendix \ref{app:obs} of this work]{Roederer2014}. More promising for the immediate future is to compare the high abundance tail of the [Mg/Fe] distributions in the simulations with observations, with the goal to constrain the SFR of the birth environments of metal poor stars. 

In order to make a qualitative comparison between all the simulated models and the observations, we first make a cut in all data sets where they overlap in $\langle [ \mathrm{Fe} / \mathrm{H}] \rangle$, which leaves us with simulated gas with $-2.6 < \langle [\mathrm{Fe} / \mathrm{H}] \rangle \leq -2.0$ (for the observations this range applies to the observed metallicity). Within this metallicity range, we identify the [Mg/Fe] abundance value below which 95\% of the stars are contained, which gives [Mg/Fe] $\simeq$ 0.45. We then calculate for each simulation the fractional mass of cold, dense gas that has an abundance higher than this threshold: MW progenitor high SFR 21\%, MW progenitor low SFR 13\% and Satellite 15\%. This begins to pose a problem for the high SFR model. Furthermore, if the SNe yields are adjusted to be compatible with the stars at lowest metallicity (where models are not yet in steady state, but the stars show a larger dispersion in [Mg/Fe]), the high SFR model is even less capable of explaining the observations. This is because, the rate of smoothing down of the yield variance depends sensitively on the turbulent diffusion coefficient which is significantly lower for the high SFR model (see Section \ref{sec:discussion}). Obviously, the above comparison is only sketchy and should be taken as an order of magnitude estimate at present given the challenges discussed above.

It is evident from Figure~\ref{fig:cold_dense_abundances} that the SFR leaves an imprint on the metal spread, whose amplitude at a given mean metallicity is primarily influenced by SN-driven turbulence, whose coefficient determines the speed at which the initial [Mg/Fe] spread is smoothed out as the $\langle [ \mathrm{Fe} / \mathrm{H}] \rangle$ increases in the simulations (Table \ref{tab:simulations}). From the simple analysis presented above, we find evidence that the properties of the metal poor halo stars are consistent with stellar birth sites in a MW progenitor with a low SFR or a satellite galaxy.

Our findings provide ancillary evidence that the dispersion of [Mg/Fe] observed in metal poor stars are consistent with either these stars being deposited into the stellar halo of the MW by mergers of low mass dwarf galaxies \citep[e.g.,][]{Bullock2001, Robertson2005, Johnston2008} or potentially in-situ formation if the MW had a modest SFR, as has been recently suggested by \citet{Wang2021}.

\section{discussion}
\label{sec:discussion}
Motions in a turbulent flow endure over a broad range of length and time scales. The characteristic length scales correspond to the fluctuating eddy motions that exist, with the largest scales bounded by the geometric dimensions of the flow. The diffusion coefficient associated with turbulence can be estimated as \citep{Karlsson2013, Krumholz2018}
\begin{equation}
\kappa \approx \frac{z_{\rm eff} \langle \sigma_{\rm v} \rangle}{3},
\end{equation}
where $z_{\rm eff}$ is the outer scale of the turbulence (scale height of the disk) and $\langle \sigma_{v} \rangle$ is the gas turbulent velocity dispersion on this scale.
Table~\ref{tab:simulations} gives the diffusion coefficient for all the models presented in this work. Higher diffusion coefficients indicate faster rates of mixing. For comparison, in the solar neighborhood $\kappa \approx 300$\,pc\,km\,s$^{-1}$ \citep{Krumholz2018}. We note that 
the diffusion coefficient derived here for dwarf galaxy environments, which is larger than the ones derived for MW progenitors, is concordant with the sub-grid diffusion models that are needed in cosmological simulations in order to explain the observed stellar [Fe/H] abundance distributions in dwarf galaxies in the Local Group \citep{2018MNRAS.474.2194E}.

An important quality of metal mixing is the connection that exists between the variance of the spatial metal concentration to the diffusion coefficient. \citet{Krumholz2018} used a stochastically forced diffusion framework to model the spatial metal fluctuations in galaxies and derive a simple estimate for a characteristic mixing length scale 
\begin{equation}
x_{\rm m} \approx \left( \frac{\kappa}{\Gamma}\right)^{1/4},
\label{eq:xm}
\end{equation}
where $\Gamma$ is the rate of SN events per unit area. Using the values for $\kappa$ and $\Gamma$ listed in Table~\ref{tab:simulations} along with Equation~(\ref{eq:xm}), we derive $x_{\rm m} = 26$\,pc and 67\,pc for the high SFR and the low SFR models, respectively. This simple estimate is in close agreement with the characteristic mixing mass scales calculated in Section~\ref{sec:mix_mass} for the high SFR and the low SFR models at the fixed $\langle [\text{Fe}/\text{H}] \rangle \approx -2.3$: $x_{\rm mix}=\SI{31}{pc}$ and $x_{\rm mix}=\SI{78}{pc}$, respectively. As we argue in Section \ref{sec:mixing}, the SFR plays an important role in determining the scale of the metal abundance fluctuations and can be inferred by studying the metallicity dispersion in stars.\footnote{We note here that the typical abundance scatter predicted by \citet{Krumholz2018} is smaller than the one deduced from our simulations and observed in metal poor stars. This is to be expected, as their model is manufactured to predict abundance dispersions for the MW at solar metallicity.}\enlargethispage{\baselineskip}

The mixing mass scale measures a characteristic enlargement of the effective size of a remnant. At this scale, the metallicity fluctuations start to be smoothed out. An akin pressure length scale \citep{Draine2011} can be estimated by calculating the characteristic radius at which the SN blast wave expansion velocity becomes comparable to $\langle \sigma_{v} \rangle$:
\begin{equation}
 x_{P}\approx 67\; \rho_{0,-24}^{-0.37} \left( \frac{E_{\rm SN}}{ 10^{51}{\rm erg}} \right)^{0.32} \left( \frac{\langle \sigma_{v} \rangle}{10{\rm km/s}} \right)^{-0.4} {\rm pc},
\end{equation}
where $\rho_{0, -24}=\frac{\rho_0}{10^{-24}\;{\rm g\;cm^{-3}}}$. Taking $\rho_0$ as the mid plane density in the galactic disk yields $x_{P} = \SI{20}{pc}$ and $x_{P} = \SI{59}{pc}$ for the high SFR and the low SFR models, respectively. We caution the reader that these length scales do not take into account galactic winds \citep[e.g.,][]{Christensen2018} and that $x_{P}$ and $x_{\rm mix}$ are comparable to the scale height of the disk. In Section~\ref{sec:mix_mass}, we show that a sizable number of remnants are advected by the galactic wind before they are able to effectively mix with the ISM. 
This implies that the winds simply remove portions of the gas that are not widely distributed and are highly correlated with the metal field. Therefore, the effect of metal removal on the dispersion of the metal field is limited. Similarly, we expect galactic shear \citep{Yang2012, Petit2015} only to be prominent on relatively longer timescales when compared with the timescale for mixing of gas by turbulent feedback from SNe \citep[e.g.,][]{Pan2010, Pan2013}.

Observations of [Mg/Fe] dispersion in metal poor stars suggest variations in the metal yields of SNe. Such intrinsic variations are preserved when the ISM is not well mixed on average at these early times. They are subsequently smoothed out by turbulent diffusion. As we have argued here, the degree of smoothing is sensitive to the SFR. The observed dispersion values in stellar abundances of stars in the Milky Way can be explained when considering the early spotty pollution of new elements from non-chemically homogeneous SNe \citep{AudouzeSilk1995}. For a hierarchically structured ISM with SF operating on a local dynamical timescale, the size of OB stellar associations is typically comparable with the metal mixing scale \citep[e.g.,][]{1998ASPC..147..278E} and the new elements should continue to be patchy and variable for stars formed in different clusters. That is, such gas variations are expected to be preserved by the SF process.

The statistics of variations of stars in chemical space can provide valuable constraints on the star forming conditions. According to our study, the observed abundance spread of [Mg/Fe] in metal poor halo stars \citep{Roederer2014} is consistent with the SF properties of 
dwarf satellites or an early low SFR MW like the one described by \citet{Wang2021}. Our findings are broadly consistent with the idea that the metal poor stellar content of the MW halo could in principle be explained entirely by disrupted dwarf satellites \citep[e.g.,][]{Bullock2001,Robertson2005, Johnston2008, Frebel2010,Roederer2018}, which has recently gained stellar kinematic support \citep{Santistevan2021}. 

Metal variations in halo stars thus represent an opportunity for chemical tagging studies to further elucidate their formation origins. To better understand the stellar halo and its assembly, it is crucial to select halo stars without introducing any biases, while reducing the contamination introduce by combining multiple stellar populations. In principle, large spectroscopic surveys of MW stars such as Gaia-ESO \citep[][]{GaiaESOsurvey} offer a unique opportunity for reconstructing the SFH of the MW by measuring the metal abundance spreads in the Galactic field and, in particular, among stars residing in individual stellar streams. 

Disentangling the SFHs of the many satellites that contributed to stars in phase-mixed galactic fields is challenging, as illustrated in Section~\ref{sec:stars}. More optimistically, we hope that for stars residing in individual streams, metal variations may be key to understand the SFH of the progenitor galaxy. For surviving galaxies, it might be possible to deduce if there are clear differences between the formation history of stars belonging to the surviving self-bound satellite and the debris stars of the disrupted accretion event.

Our understanding of the assembly of the MW has come a long way. Nevertheless, it continues to offer major puzzles and challenges. Metal variations thus provide us with an exciting opportunity to derive, in combination with kinematic studies and stellar tagging, the contribution of satellites to the accreted stellar halo mass, as well as, inferring the specific SFHs of individual galaxies.

\begin{acknowledgements}
We thank the referee for insightful questions that have helped to substantially improve the quality of the paper. We thank S.\,Faber, E.\,Kirby, A.\,Ji, K.\,Brauer, P.\,Macias, J.\,Primack and D.\,Zaritsky, for useful discussions. A.N.K. and E.R.-R. acknowledge support by Heising-Simons Foundation, the Danish National Research Foundation (DNRF132) and NSF (AST-1911206 and AST-1852393). This work made use of an HPC facility funded by a grant from VILLUM FONDEN (project number 16599). H.P. acknowledges support from the Danish National Research Foundation (DNRF132) and the Hong Kong government (GRF grant HKU27305119, HKU17304821).
M.S.F gratefully acknowledges support provided by NASA through Hubble Fellowship grant HST-HF2-51493.001-A awarded by the Space Telescope Science Institute, which is operated by the Association of Universities for Research in Astronomy, In., for NASA, under the contract NAS 5-26555. 
\end{acknowledgements}

\bibliographystyle{aasjournal}
\bibliography{references}

\begin{thebibliography}{}
\expandafter\ifx\csname natexlab\endcsname\relax\def\natexlab#1{#1}\fi
\providecommand{\url}[1]{\href{#1}{#1}}
\providecommand{\dodoi}[1]{doi:~\href{http://doi.org/#1}{\nolinkurl{#1}}}
\providecommand{\doeprint}[1]{\href{http://ascl.net/#1}{\nolinkurl{http://ascl.net/#1}}}
\providecommand{\doarXiv}[1]{\href{https://arxiv.org/abs/#1}{\nolinkurl{https://arxiv.org/abs/#1}}}

\bibitem[{{Abohalima} \& {Frebel}(2018)}]{Jinabase}
{Abohalima}, A., \& {Frebel}, A. 2018, \apjs, 238, 36,
  \dodoi{10.3847/1538-4365/aadfe9}

\bibitem[{{Aoki} {et~al.}(2007){Aoki}, {Beers}, {Christlieb}, {Norris}, {Ryan},
  \& {Tsangarides}}]{Aoki2007}
{Aoki}, W., {Beers}, T.~C., {Christlieb}, N., {et~al.} 2007, \apj, 655, 492,
  \dodoi{10.1086/509817}

\bibitem[{{Armillotta} {et~al.}(2018){Armillotta}, {Krumholz}, \&
  {Fujimoto}}]{Armillotta2018}
{Armillotta}, L., {Krumholz}, M.~R., \& {Fujimoto}, Y. 2018, \mnras, 481, 5000,
  \dodoi{10.1093/mnras/sty2625}

\bibitem[{{Arnone} {et~al.}(2005){Arnone}, {Ryan}, {Argast}, {Norris}, \&
  {Beers}}]{Arnone2005}
{Arnone}, E., {Ryan}, S.~G., {Argast}, D., {Norris}, J.~E., \& {Beers}, T.~C.
  2005, \aap, 430, 507, \dodoi{10.1051/0004-6361:20041034}

\bibitem[{{Audouze} \& {Silk}(1995)}]{AudouzeSilk1995}
{Audouze}, J., \& {Silk}, J. 1995, \apjl, 451, L49, \dodoi{10.1086/309687}

\bibitem[{{Beers} \& {Christlieb}(2005)}]{Beers2005}
{Beers}, T.~C., \& {Christlieb}, N. 2005, \araa, 43, 531,
  \dodoi{10.1146/annurev.astro.42.053102.134057}

\bibitem[{{Bullock} {et~al.}(2001){Bullock}, {Kravtsov}, \&
  {Weinberg}}]{Bullock2001}
{Bullock}, J.~S., {Kravtsov}, A.~V., \& {Weinberg}, D.~H. 2001, \apj, 548, 33,
  \dodoi{10.1086/318681}

\bibitem[{{Cayrel} {et~al.}(2004){Cayrel}, {Depagne}, {Spite}, {Hill}, {Spite},
  {Fran{\c{c}}ois}, {Plez}, {Beers}, {Primas}, {Andersen}, {Barbuy},
  {Bonifacio}, {Molaro}, \& {Nordstr{\"o}m}}]{Cayrel2004}
{Cayrel}, R., {Depagne}, E., {Spite}, M., {et~al.} 2004, \aap, 416, 1117,
  \dodoi{10.1051/0004-6361:20034074}

\bibitem[{Christensen {et~al.}(2018)Christensen, Dave, Brooks, Quinn, \&
  Shen}]{Christensen2018}
Christensen, C.~R., Dave, R., Brooks, A., Quinn, T., \& Shen, S. 2018,
  \dodoi{10.3847/1538-4357/aae374}

\bibitem[{{Colbrook} {et~al.}(2017){Colbrook}, {Ma}, {Hopkins}, \&
  {Squire}}]{Colbrook2017}
{Colbrook}, M.~J., {Ma}, X., {Hopkins}, P.~F., \& {Squire}, J. 2017, \mnras,
  467, 2421, \dodoi{10.1093/mnras/stx261}

\bibitem[{{Dalcanton}(2007)}]{Dalcanton2007}
{Dalcanton}, J.~J. 2007, \apj, 658, 941, \dodoi{10.1086/508913}

\bibitem[{{de Avillez} \& {Mac Low}(2002)}]{Avillez2002}
{de Avillez}, M.~A., \& {Mac Low}, M.-M. 2002, \apj, 581, 1047,
  \dodoi{10.1086/344256}

\bibitem[{{Deason} {et~al.}(2016){Deason}, {Mao}, \& {Wechsler}}]{Deason2016}
{Deason}, A.~J., {Mao}, Y.-Y., \& {Wechsler}, R.~H. 2016, \apj, 821, 5,
  \dodoi{10.3847/0004-637X/821/1/5}

\bibitem[{{Draine}(2011)}]{Draine2011}
{Draine}, B.~T. 2011, {Physics of the Interstellar and Intergalactic Medium}

\bibitem[{{Eggen} {et~al.}(1962){Eggen}, {Lynden-Bell}, \&
  {Sandage}}]{Eggen1962}
{Eggen}, O.~J., {Lynden-Bell}, D., \& {Sandage}, A.~R. 1962, \apj, 136, 748,
  \dodoi{10.1086/147433}

\bibitem[{{Elmegreen}(1998)}]{1998ASPC..147..278E}
{Elmegreen}, B.~G. 1998, in Astronomical Society of the Pacific Conference
  Series, Vol. 147, Abundance Profiles: Diagnostic Tools for Galaxy History,
  ed. D.~{Friedli}, M.~{Edmunds}, C.~{Robert}, \& L.~{Drissen}, 278.
\newblock \doarXiv{astro-ph/9712354}

\bibitem[{{Escala} {et~al.}(2018){Escala}, {Wetzel}, {Kirby}, {Hopkins}, {Ma},
  {Wheeler}, {Kere{\v{s}}}, {Faucher-Gigu{\`e}re}, \&
  {Quataert}}]{2018MNRAS.474.2194E}
{Escala}, I., {Wetzel}, A., {Kirby}, E.~N., {et~al.} 2018, \mnras, 474, 2194,
  \dodoi{10.1093/mnras/stx2858}

\bibitem[{{Few} {et~al.}(2014){Few}, {Courty}, {Gibson}, {Michel-Dansac}, \&
  {Calura}}]{Few2014}
{Few}, C.~G., {Courty}, S., {Gibson}, B.~K., {Michel-Dansac}, L., \& {Calura},
  F. 2014, \mnras, 444, 3845, \dodoi{10.1093/mnras/stu1709}

\bibitem[{Fielding {et~al.}(2017)Fielding, Quataert, Martizzi, \&
  Faucher-Giguere}]{Fielding2017}
Fielding, D., Quataert, E., Martizzi, D., \& Faucher-Giguere, C.-A. 2017, 5, 1,
  \dodoi{10.1093/mnrasl/slx072}

\bibitem[{Fragile {et~al.}(2003)Fragile, Murray, Anninos, \& Lin}]{Fragile2003}
Fragile, P.~C., Murray, S.~D., Anninos, P., \& Lin, D. N.~C. 2003, The
  Astrophysical Journal, 590, 778–790, \dodoi{10.1086/375183}

\bibitem[{{Frebel}(2010)}]{Frebel2010}
{Frebel}, A. 2010, Astronomische Nachrichten, 331, 474,
  \dodoi{10.1002/asna.201011362}

\bibitem[{Frebel \& Norris(2015)}]{Frebel2015}
Frebel, A., \& Norris, J.~E. 2015, Annual Review of Astronomy and Astrophysics,
  53, 631, \dodoi{10.1146/annurev-astro-082214-122423}

\bibitem[{{Gilmore} {et~al.}(2012){Gilmore}, {Randich}, {Asplund}, {Binney},
  {Bonifacio}, {Drew}, {Feltzing}, {Ferguson}, {Jeffries}, {Micela},
  {Negueruela}, {Prusti}, {Rix}, {Vallenari}, {Alfaro}, {Allende-Prieto},
  {Babusiaux}, {Bensby}, {Blomme}, {Bragaglia}, {Flaccomio}, {Fran{\c{c}}ois},
  {Irwin}, {Koposov}, {Korn}, {Lanzafame}, {Pancino}, {Paunzen},
  {Recio-Blanco}, {Sacco}, {Smiljanic}, {Van Eck}, {Walton}, {Aden}, {Aerts},
  {Affer}, {Alcala}, {Altavilla}, {Alves}, {Antoja}, {Arenou}, {Argiroffi},
  {Asensio Ramos}, {Bailer-Jones}, {Balaguer-Nunez}, {Bayo}, {Barbuy},
  {Barisevicius}, {Barrado y Navascues}, {Battistini}, {Bellas Velidis},
  {Bellazzini}, {Belokurov}, {Bergemann}, {Bertelli}, {Biazzo}, {Bienayme},
  {Bland-Hawthorn}, {Boeche}, {Bonito}, {Boudreault}, {Bouvier}, {Brandao},
  {Brown}, {de Bruijne}, {Burleigh}, {Caballero}, {Caffau}, {Calura},
  {Capuzzo-Dolcetta}, {Caramazza}, {Carraro}, {Casagrande}, {Casewell},
  {Chapman}, {Chiappini}, {Chorniy}, {Christlieb}, {Cignoni}, {Cocozza},
  {Colless}, {Collet}, {Collins}, {Correnti}, {Covino}, {Crnojevic}, {Cropper},
  {Cunha}, {Damiani}, {David}, {Delgado}, {Duffau}, {Edvardsson}, {Eldridge},
  {Enke}, {Eriksson}, {Evans}, {Eyer}, {Famaey}, {Fellhauer}, {Ferreras},
  {Figueras}, {Fiorentino}, {Flynn}, {Folha}, {Franciosini}, {Frasca},
  {Freeman}, {Fremat}, {Friel}, {Gaensicke}, {Gameiro}, {Garzon}, {Geier},
  {Geisler}, {Gerhard}, {Gibson}, {Gomboc}, {Gomez}, {Gonzalez-Fernandez},
  {Gonzalez Hernandez}, {Gosset}, {Grebel}, {Greimel}, {Groenewegen},
  {Grundahl}, {Guarcello}, {Gustafsson}, {Hadrava}, {Hatzidimitriou}, {Hambly},
  {Hammersley}, {Hansen}, {Haywood}, {Heber}, {Heiter}, {Held}, {Helmi},
  {Hensler}, {Herrero}, {Hill}, {Hodgkin}, {Huelamo}, {Huxor}, {Ibata},
  {Jackson}, {de Jong}, {Jonker}, {Jordan}, {Jordi}, {Jorissen}, {Katz},
  {Kawata}, {Keller}, {Kharchenko}, {Klement}, {Klutsch}, {Knude}, {Koch},
  {Kochukhov}, {Kontizas}, {Koubsky}, {Lallement}, {de Laverny}, {van Leeuwen},
  {Lemasle}, {Lewis}, {Lind}, {Lindstrom}, {Lobel}, {Lopez Santiago}, {Lucas},
  {Ludwig}, {Lueftinger}, {Magrini}, {Maiz Apellaniz}, {Maldonado}, {Marconi},
  {Marino}, {Martayan}, {Martinez-Valpuesta}, {Matijevic}, {McMahon},
  {Messina}, {Meyer}, {Miglio}, {Mikolaitis}, {Minchev}, {Minniti}, {Moitinho},
  {Momany}, {Monaco}, {Montalto}, {Monteiro}, {Monier}, {Montes}, {Mora},
  {Moraux}, {Morel}, {Mowlavi}, {Mucciarelli}, {Munari}, {Napiwotzki},
  {Nardetto}, {Naylor}, {Naze}, {Nelemans}, {Okamoto}, {Ortolani}, {Pace},
  {Palla}, {Palous}, {Parker}, {Penarrubia}, {Pillitteri}, {Piotto}, {Posbic},
  {Prisinzano}, {Puzeras}, {Quirrenbach}, {Ragaini}, {Read}, {Read}, {Reyle},
  {De Ridder}, {Robichon}, {Robin}, {Roeser}, {Romano}, {Royer}, {Ruchti},
  {Ruzicka}, {Ryan}, {Ryde}, {Santos}, {Sanz Forcada}, {Sarro Baro},
  {Sbordone}, {Schilbach}, {Schmeja}, {Schnurr}, {Schoenrich}, {Scholz},
  {Seabroke}, {Sharma}, {De Silva}, {Smith}, {Solano}, {Sordo}, {Soubiran},
  {Sousa}, {Spagna}, {Steffen}, {Steinmetz}, {Stelzer}, {Stempels},
  {Tabernero}, {Tautvaisiene}, {Thevenin}, {Torra}, {Tosi}, {Tolstoy}, {Turon},
  {Walker}, {Wambsganss}, {Worley}, {Venn}, {Vink}, {Wyse}, {Zaggia},
  {Zeilinger}, {Zoccali}, {Zorec}, {Zucker}, {Zwitter}, \& {Gaia-ESO Survey
  Team}}]{GaiaESOsurvey}
{Gilmore}, G., {Randich}, S., {Asplund}, M., {et~al.} 2012, The Messenger, 147,
  25

\bibitem[{{Johnston} {et~al.}(2008){Johnston}, {Bullock}, {Sharma}, {Font},
  {Robertson}, \& {Leitner}}]{Johnston2008}
{Johnston}, K.~V., {Bullock}, J.~S., {Sharma}, S., {et~al.} 2008, \apj, 689,
  936, \dodoi{10.1086/592228}

\bibitem[{{Karlsson} {et~al.}(2013){Karlsson}, {Bromm}, \&
  {Bland-Hawthorn}}]{Karlsson2013}
{Karlsson}, T., {Bromm}, V., \& {Bland-Hawthorn}, J. 2013, Reviews of Modern
  Physics, 85, 809, \dodoi{10.1103/RevModPhys.85.809}

\bibitem[{{Karpov} {et~al.}(2020){Karpov}, {Martizzi}, {Macias},
  {Ramirez-Ruiz}, {Kolborg}, \& {Naiman}}]{Karpov2020}
{Karpov}, P.~I., {Martizzi}, D., {Macias}, P., {et~al.} 2020, \apj, 896, 66,
  \dodoi{10.3847/1538-4357/ab8f23}

\bibitem[{Kobayashi {et~al.}(2006)Kobayashi, Umeda, Nomoto, Tominaga, \&
  Ohkubo}]{Kobayashi2006}
Kobayashi, C., Umeda, H., Nomoto, K., Tominaga, N., \& Ohkubo, T. 2006, The
  Astrophysical Journal, 653, 1145, \dodoi{10.1086/508914}

\bibitem[{Krumholz \& Ting(2018)}]{Krumholz2018}
Krumholz, M.~R., \& Ting, Y.-s. 2018, Monthly Notices of the Royal Astronomical
  Society, 475, 2236, \dodoi{10.1093/mnras/stx3286}

\bibitem[{{Mac Low} \& {Ferrara}(1999)}]{MacLow1999}
{Mac Low}, M.-M., \& {Ferrara}, A. 1999, \apj, 513, 142, \dodoi{10.1086/306832}

\bibitem[{Macias \& Ramirez-Ruiz(2018)}]{Macias2018}
Macias, P., \& Ramirez-Ruiz, E. 2018, The Astrophysical Journal, 860, 89,
  \dodoi{10.3847/1538-4357/aac3e0}

\bibitem[{{Mackey} {et~al.}(2003){Mackey}, {Bromm}, \&
  {Hernquist}}]{2003ApJ...586....1M}
{Mackey}, J., {Bromm}, V., \& {Hernquist}, L. 2003, \apj, 586, 1,
  \dodoi{10.1086/367613}

\bibitem[{Martizzi {et~al.}(2015)Martizzi, Faucher-Gigu{\`{e}}re, \&
  Quataert}]{Martizzi2015}
Martizzi, D., Faucher-Gigu{\`{e}}re, C.~A., \& Quataert, E. 2015, Monthly
  Notices of the Royal Astronomical Society, 450, 504,
  \dodoi{10.1093/mnras/stv562}

\bibitem[{Martizzi {et~al.}(2016)Martizzi, Fielding, Faucher-Gigu{\`{e}}re, \&
  Quataert}]{Martizzi2016}
Martizzi, D., Fielding, D., Faucher-Gigu{\`{e}}re, C.~A., \& Quataert, E. 2016,
  Monthly Notices of the Royal Astronomical Society, 459, 2311,
  \dodoi{10.1093/mnras/stw745}

\bibitem[{{Naiman} {et~al.}(2018){Naiman}, {Pillepich}, {Springel},
  {Ramirez-Ruiz}, {Torrey}, {Vogelsberger}, {Pakmor}, {Nelson}, {Marinacci},
  {Hernquist}, {Weinberger}, \& {Genel}}]{Naiman2018}
{Naiman}, J.~P., {Pillepich}, A., {Springel}, V., {et~al.} 2018, \mnras, 477,
  1206, \dodoi{10.1093/mnras/sty618}

\bibitem[{Nomoto {et~al.}(2013)Nomoto, Kobayashi, \& Tominaga}]{Nomoto2013}
Nomoto, K., Kobayashi, C., \& Tominaga, N. 2013, Annual Review of Astronomy and
  Astrophysics, 51, 457, \dodoi{10.1146/annurev-astro-082812-140956}

\bibitem[{{Pan} \& {Scannapieco}(2010)}]{Pan2010}
{Pan}, L., \& {Scannapieco}, E. 2010, \apj, 721, 1765,
  \dodoi{10.1088/0004-637X/721/2/1765}

\bibitem[{Pan {et~al.}(2013)Pan, Scannapieco, \& Scalo}]{Pan2013}
Pan, L., Scannapieco, E., \& Scalo, J. 2013, The Astrophysical Journal, 775,
  34, \dodoi{10.1088/0004-637X/775/2/111}

\bibitem[{Petit {et~al.}(2015)Petit, Krumholz, Goldbaum, \& Forbes}]{Petit2015}
Petit, A.~C., Krumholz, M.~R., Goldbaum, N.~J., \& Forbes, J.~C. 2015, Monthly
  Notices of the Royal Astronomical Society, 449, 2588,
  \dodoi{10.1093/mnras/stv493}

\bibitem[{{Robertson} {et~al.}(2005){Robertson}, {Bullock}, {Font}, {Johnston},
  \& {Hernquist}}]{Robertson2005}
{Robertson}, B., {Bullock}, J.~S., {Font}, A.~S., {Johnston}, K.~V., \&
  {Hernquist}, L. 2005, \apj, 632, 872, \dodoi{10.1086/452619}

\bibitem[{{Roederer} {et~al.}(2018){Roederer}, {Hattori}, \&
  {Valluri}}]{Roederer2018}
{Roederer}, I.~U., {Hattori}, K., \& {Valluri}, M. 2018, \aj, 156, 179,
  \dodoi{10.3847/1538-3881/aadd9c}

\bibitem[{Roederer {et~al.}(2014)Roederer, Preston, Thompson, Shectman, Sneden,
  Burley, \& Kelson}]{Roederer2014}
Roederer, I.~U., Preston, G.~W., Thompson, I.~B., {et~al.} 2014, Astronomical
  Journal, 147, \dodoi{10.1088/0004-6256/147/6/136}

\bibitem[{{Ryan} {et~al.}(1996){Ryan}, {Norris}, \& {Beers}}]{Ryan1996}
{Ryan}, S.~G., {Norris}, J.~E., \& {Beers}, T.~C. 1996, \apj, 471, 254,
  \dodoi{10.1086/177967}

\bibitem[{{Santistevan} {et~al.}(2021){Santistevan}, {Wetzel}, {Sanderson},
  {El-Badry}, {Samuel}, \& {Faucher-Gigu{\`e}re}}]{Santistevan2021}
{Santistevan}, I.~B., {Wetzel}, A., {Sanderson}, R.~E., {et~al.} 2021, \mnras,
  505, 921, \dodoi{10.1093/mnras/stab1345}

\bibitem[{{Searle} \& {Zinn}(1978)}]{Searle1978}
{Searle}, L., \& {Zinn}, R. 1978, \apj, 225, 357, \dodoi{10.1086/156499}

\bibitem[{Shen {et~al.}(2015)Shen, Cooke, Ramirez-Ruiz, Madau, Mayer, \&
  Guedes}]{Shen2015}
Shen, S., Cooke, R.~J., Ramirez-Ruiz, E., {et~al.} 2015, Astrophysical Journal,
  807, 115, \dodoi{10.1088/0004-637X/807/2/115}

\bibitem[{{Shen} {et~al.}(2017){Shen}, {Kulkarni}, {Madau}, \&
  {Mayer}}]{Shen2017}
{Shen}, S., {Kulkarni}, G., {Madau}, P., \& {Mayer}, L. 2017, \mnras, 469,
  4012, \dodoi{10.1093/mnras/stx1094}

\bibitem[{Teyssier(2002)}]{Teyssier2002}
Teyssier, R. 2002, Astronomy {\&} Astrophysics, 385, 337

\bibitem[{{van Dokkum} {et~al.}(2013){van Dokkum}, {Leja}, {Nelson}, {Patel},
  {Skelton}, {Momcheva}, {Brammer}, {Whitaker}, {Lundgren}, {Fumagalli},
  {Conroy}, {F{\"o}rster Schreiber}, {Franx}, {Kriek}, {Labb{\'e}},
  {Marchesini}, {Rix}, {van der Wel}, \& {Wuyts}}]{vanDokkum2013}
{van Dokkum}, P.~G., {Leja}, J., {Nelson}, E.~J., {et~al.} 2013, \apjl, 771,
  L35, \dodoi{10.1088/2041-8205/771/2/L35}

\bibitem[{{Wang} {et~al.}(2021){Wang}, {Nadler}, {Mao}, {Adhikari}, {Wechsler},
  \& {Behroozi}}]{Wang2021}
{Wang}, Y., {Nadler}, E.~O., {Mao}, Y.-Y., {et~al.} 2021, arXiv e-prints,
  arXiv:2102.11876.
\newblock \doarXiv{2102.11876}

\bibitem[{{Welsh} {et~al.}(2021){Welsh}, {Cooke}, \& {Fumagalli}}]{Welsh2021}
{Welsh}, L., {Cooke}, R., \& {Fumagalli}, M. 2021, \mnras, 500, 5214,
  \dodoi{10.1093/mnras/staa3342}

\bibitem[{Yang \& Krumholz(2012)}]{Yang2012}
Yang, C.~C., \& Krumholz, M. 2012, Astrophysical Journal, 758, 1,
  \dodoi{10.1088/0004-637X/758/1/48}

\end{thebibliography}

\appendix
\renewcommand\thefigure{\thesection.\arabic{figure}} 
\setcounter{figure}{0} 

\section{Observed abundances and SN yields}
\label{app:obs}

\begin{figure}
    \centering
    \includegraphics[width = 0.8\textwidth]{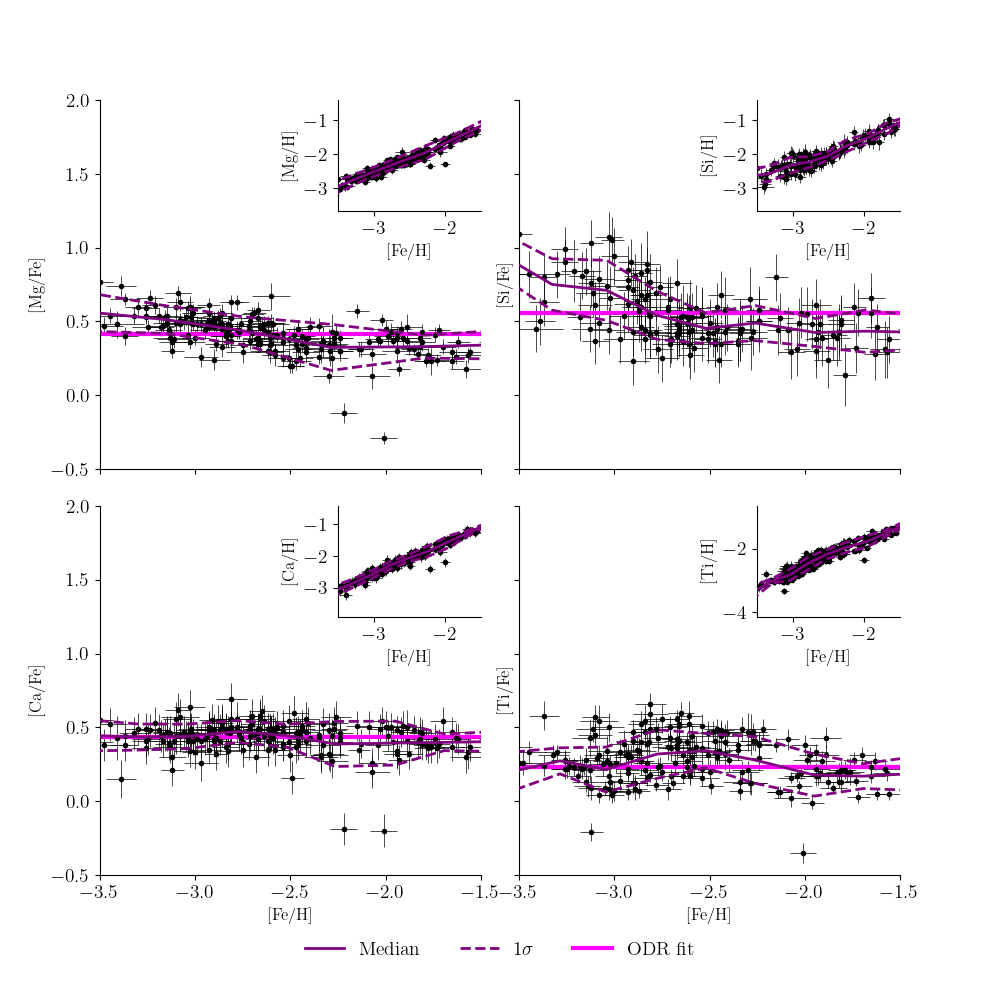}
    \caption{Abundances of selected $\alpha$-elements of the metal poor halo stars from \citet{Roederer2014} which we use as our comparison sample. From top left, to bottom right, the panels show the measured abundances of: Mg, Si, Ca and Ti relative to Fe as function of [Fe/H]. Error bars indicate the total uncertainty as reported by \citet{Roederer2014}. The solid, dark purple lines indicate the mean in 0.5 dex wide bins in [Fe/H] and the dashed lines are the $1 \sigma$ spread in the same bins. The insets in each panel shows the abundances of the same elements relative to H. The solid fuchsia lines indicate the best orthogonal distance regression (ODR) fit to the abundances.}
    \label{fig:alpha_elemns}
\end{figure}

\begin{figure}
\begin{minipage}[t]{.475\textwidth}
  \centering
    \includegraphics[height = 5 cm]{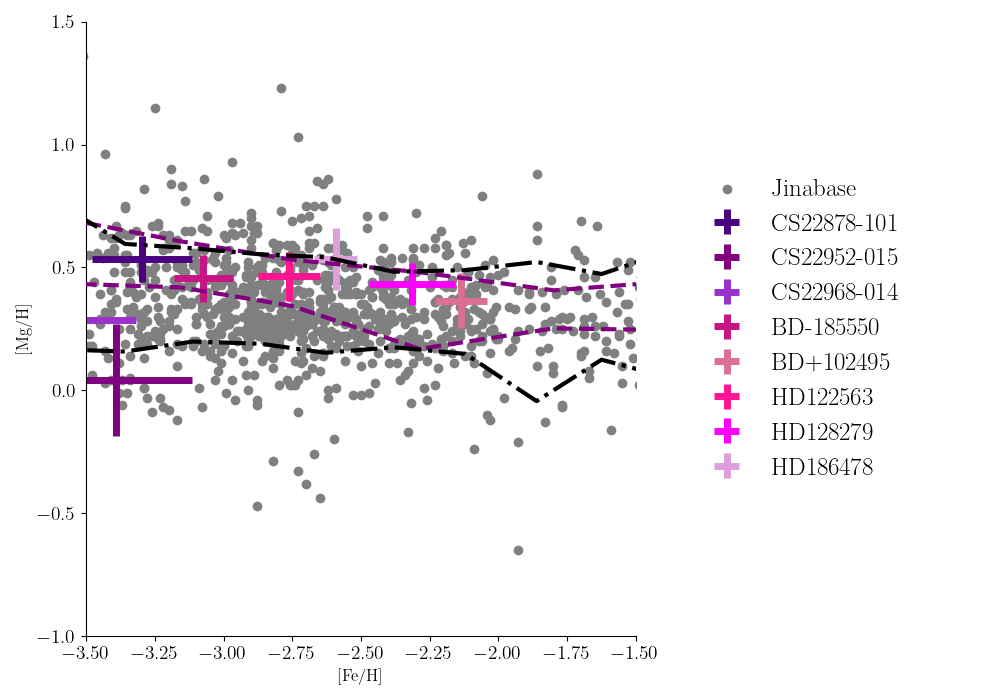}
  \caption{[Mg/Fe] versus [Fe/H] abundance measurements of non-CEMP stars in the MW halo from Jinabase \citep{Jinabase}. The colored crosses the intrinsic dispersion of measurements for a selection of stars. Black dash-dotted lines indicate the $1 \sigma$ spread of these stars in the same [Fe/H] bins as used in Figure \ref{fig:alpha_elemns}. Purple dashed lines are the same as in the [Mg/Fe] panel in  Figure~\ref{fig:alpha_elemns} and have been added to aid comparison.}
  \label{fig:jinabase}
\end{minipage}
\hspace{0.05 \textwidth}
\begin{minipage}[t]{.475\textwidth}
  \centering
  \includegraphics[height = 5 cm]{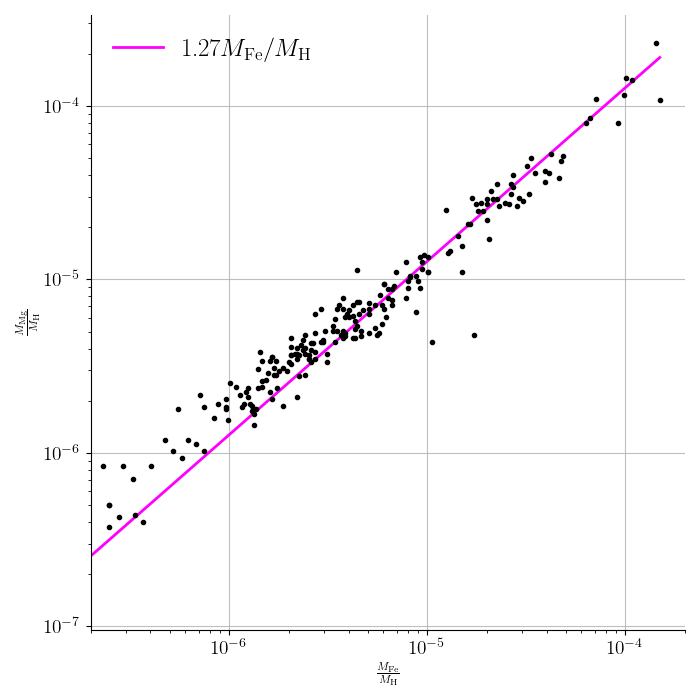}
  \caption{Mass ratio of Mg to H as a function of the same for Fe to H for the 224 stars from \citep{Roederer2014}. The solid line is the best linear fit to the data and is used to determine the mass of Fe per SNe for the chemically identical SNe models used in this project.}
  \label{fig:mass_fit}
\end{minipage}
\end{figure}

We compare our simulated results with the observations of the 313 metal poor, $-4.5 \leq [\mathrm{Fe} / \mathrm{H}] \leq -1.1$, MW halo stars deduced by \citet{Roederer2014}.  After removing stars with upper limit measurements for either [Fe/H] or [Mg/H], as well as, carbon enhanced metal poor (CEMP) stars, defined as stars with [C/Fe] $> 0.7$ \citep{Aoki2007}, we are left with 224 stars in the sample. 
CEMP stars exhibit anomalous abundances of C relative to Fe, several theories exist to explain this overabundance, one likely model is that these stars have experienced mass transfer in a binary with an asymptotic giant branch star \citep{Beers2005}. If this is the case their measured abundances will not be representative of the gas from which they formed and we thus remove them from this sample. 

In Figure \ref{fig:alpha_elemns} we show the measured abundances of the 224 stars we use from \citet{Roederer2014}, in four $\alpha$ elements: Mg, Si, Ca and Ti. We use the total uncertainty from their table 12 and the method they outline to calculate uncertainties for other ratios to get $\sigma_{[\mathrm{X}/\mathrm{H}]}$. Also plotted are the mean and $1 \sigma$ spreads calculated in 0.5 dex wide bins in [Fe/H].
We note that Ti, while not an $\alpha$-element, is the result of incomplete burning and is highly correlated with $\alpha$-element production. Baring the much larger uncertainties in Si measurements, the overall behaviour in these element abundances are very similar to the behaviours of Mg in that, they all show tight correlations with [Fe/H]  with nearly constant scatter around the mean of the range of metallicity considered here. This underscores that Mg is indeed an appropriate, representative of the $\alpha$-elements. This is to give further credence to the idea that yield variations are indeed required by the observations.

As a point of comparison we show, in the Figure \ref{fig:jinabase}, the abundances reported to Jinabase \citep{Jinabase} for MW halo stars marked as non-CEMP, fulfilling $-3.5 \leq [\mathrm{Fe}/ \mathrm{H}] \leq -1.0$. When abundances are reported several times for the same star only one data point is shown for that star, this selection was made at random. This selection results in a total of 1056 stars included. For a subset of the stars with multiple measurements we calculate the intrinsic star-to-star scatter between reported abundances from different groups, which is indicated by the colored error bars on the panel. This intrinsic scatter is indicative of the systematic errors between different analyses. The existence and size of this scatter is the reason we decided to compare our results with observations from a single study to ensure that, to the largest degree possible, the spreads reported were intrinsic rather than being dominated by systematic uncertainties between different measurements.

\subsection{SN Yields}
To determine the yield ratio of the chemically identical SNe we calculate the mass ratios of Mg to H and Fe to H of the \citep{Roederer2014} sample and fit a linear relationship. The result of this is plotted in Figure \ref{fig:mass_fit}. The best fit has slope: $\frac{M_\mathrm{Mg}}{M_\mathrm{H}} = 1.27 \frac{M_\mathrm{Fe}}{M_\mathrm{H}}$, fixing the mass of Mg per event to $0.10 M_\odot$ and requiring all SNe to obey this relationship, the mass of Fe per event needs to be $M_\mathrm{Fe} = 0.08 M_\odot$. 

\begin{figure}
    \centering
    \includegraphics[width = 0.8\textwidth]{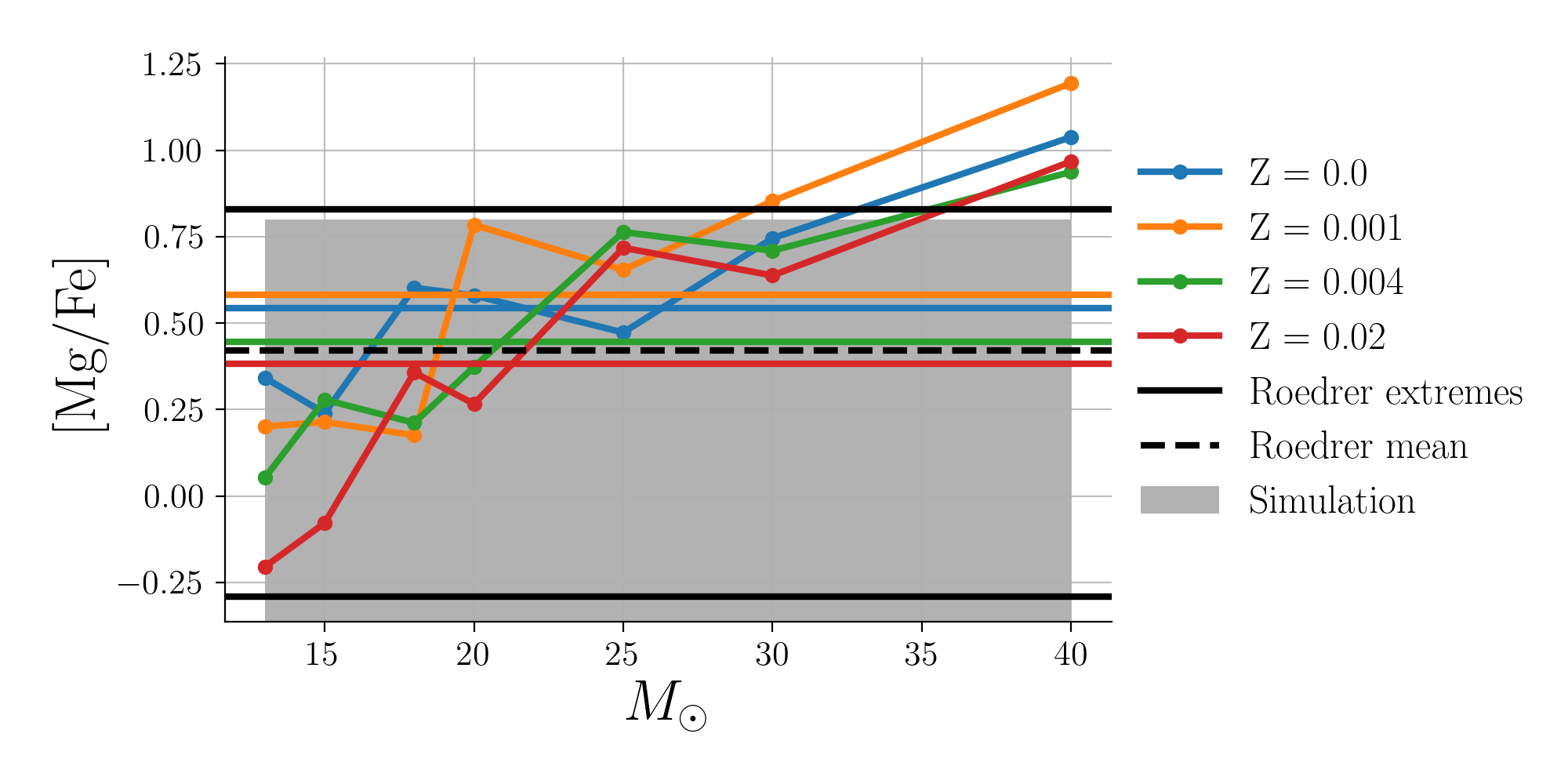}
    \caption{Yield abundances of SNe as a function of progenitor mass and metallicity according to \citep{Kobayashi2006}. The horizontal lines show the IMF weighted average yields at the same metallities. Also shown is the mean abundance of the \citep{Roederer2014} sample (solid black line) as well as the abundance of the most and least enriched stars in that sample (black dashed lines). The gray box indicates the yields used in this work [Mg/Fe] =0.72 and [Mg/Fe] = -1.39 (lower yield not visible in this panel).  }
    \label{fig:Kobayashi_yields}
\end{figure}

As described in Section \ref{sec:metal_inj} we disentangle the Mg field from the Fe by allowing SNe not to be chemically identical, as required by observations. We achieve this by introducing two specific [Mg/Fe] yields for SN injection sites. These yield values are chosen again to match the mean we expect based on the observational data set of [Mg/Fe] = 0.42. We set the  high yield  SN injection site to have a yield [Mg/Fe] = 0.72. This value is chosen based on the most highly enriched star in the dataset. The low yield SN injection site is then set to achieve the desired mean, which results in a value of [Mg/Fe] = -1.39. \\
The values of these yields as well as the range of abundances in the observational dataset are visualized in Figure \ref{fig:Kobayashi_yields}. This figure also shows the theoretically predicted yields of SN as a function of the progenitor mass and metallicity as reported by \citep{Kobayashi2006}. In addition, the IMF weighted average of these yields is also shown. It is important to note that the predicted yields of SNe by \citet{Kobayashi2006}, albeit uncertain, are consistent with the empirically derived constraints. Although the yields predicted for the highest mass stars are in contention with the observational constraints the IMF weighted average yields agree well with the data. It is important to note observations of metal poor stars and in particular the empirically derived yield  variations, can be effectively used to constrains the currently uncertain SN explosions models.

\FloatBarrier
\setcounter{figure}{0} 

\section{Exploring changes in the star formation efficiency}
\label{app:MW_eff}
\begin{figure}
 \centering
 \includegraphics[width = 0.5\textwidth]{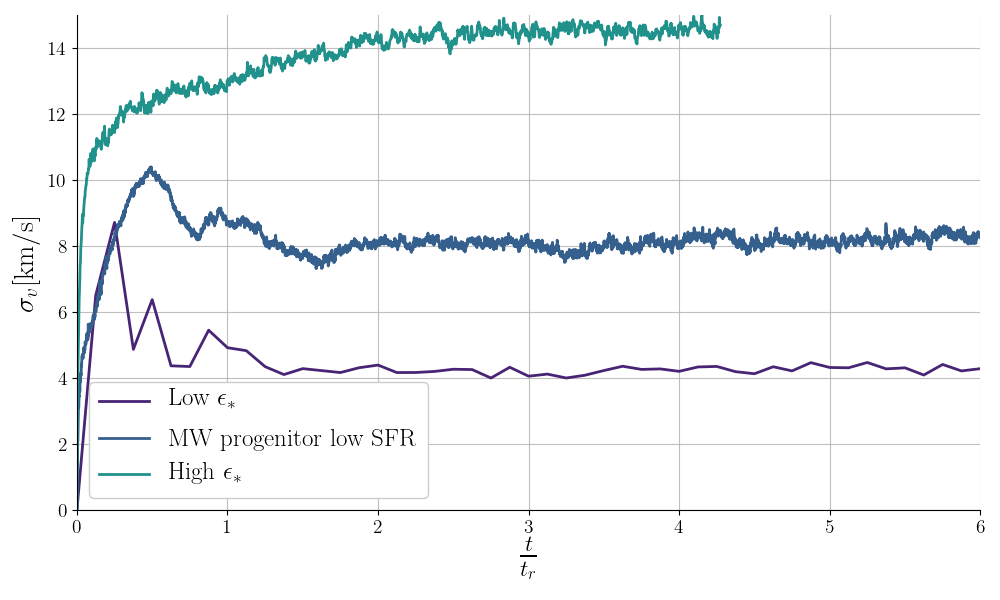}
 \caption{Comparison of the turbulent velocity dispersion, $\sigma_v$, in the three models using the same galaxy potential with varying rates of SNe per unit volume. Time is normalized by the relaxation time, $t_r$ in all cases. Both the base and low $\epsilon_*$ models reach a steady state and the velocity dispersion stays relatively constant . The high $\epsilon_*$ case, however, does not reach dynamical stability, similarly to what was observed for the satellite model in Section \ref{sec:steady}.}
 \label{fig:MW_vel_disp}
\end{figure}

Here we investigate the effects of the SF efficiency, $\epsilon_*$, in our simulations with chemically identical SNe. In particular, we explore the mixing properties of SNe occurring in three versions of the MW model with varying $\epsilon_*$. We use the MW progenitor with low SFR as our base simulation. We then increase and decrease $\epsilon_*$ by a factor of $5$, respectively. The parameters of these models are summarized in Table \ref{tab:simulations} in Section \ref{sec:sims}. In Figure \ref{fig:MW_vel_disp} we compare the time evolution of $\sigma_v$ for the three different simulations. The influence of $\epsilon_*$ on the velocity dispersion is clearly discernible. When $\epsilon_*$ is more or less extreme, there is a marked difference in the timescales for mixing of gas by the turbulence driven by SNe. $\epsilon_*$ influences the extent to which turbulence mixes and homogenizes the gas in the face of repeated SN events. Altering $\epsilon_*$ by a factor of 5 leads to a difference of $\approx 2$ in $\sigma_v$ and $t_{r}$ (Table \ref{tab:simulations}).

The value of $\sigma_v$ is set primarily by the intensity of the SN heating, which depends on the SN rate per unit volume, and the cooling rate of the shocked gas. The effect of cooling is negligible between models as they have similar gas density structures. When the energy input from SNe is equal to the binding energy of the cold gas, like in the model with highest $\epsilon_*$ and in the satellite model in Section \ref{sec:steady}, the disk becomes dynamically unstable and mass blow out occurs. Over its lifetime the high $\epsilon_*$ model loses $\approx 10 \%$ of its initial mass.

\begin{figure}
 \centering
    \includegraphics[width = \textwidth]{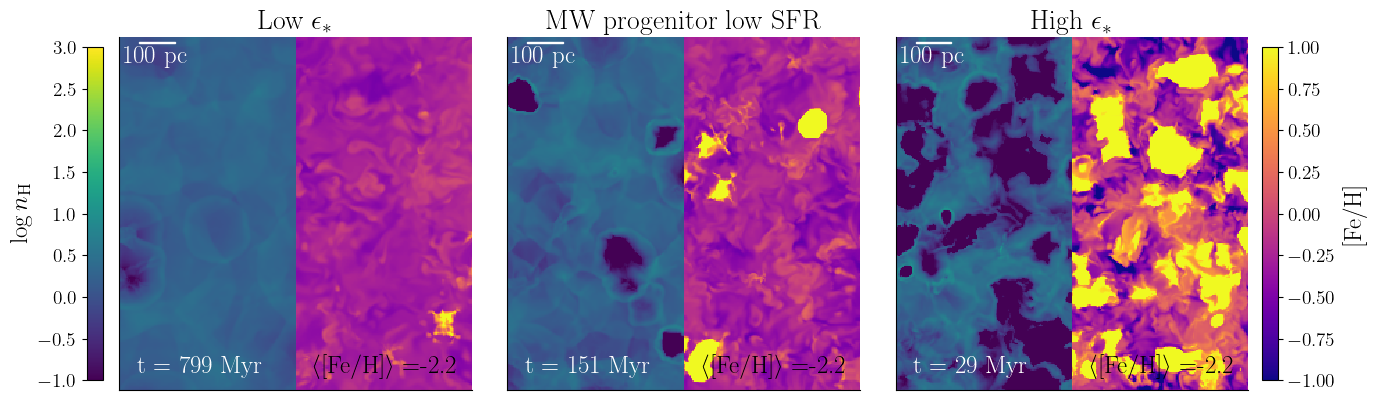}
 \caption{Same as the bottom row panels in Figure \ref{fig:maps} but for the three MW progenitor models with varying $\epsilon_*$. The slices are taken at the same mean metallicity as those in Figure \ref{fig:maps}. Because the total gas mass is the same in the three models the time to reach the same level of enrichment depends solely on $\epsilon_*$. The effects of rapid enrichment for the high $\epsilon_*$ model are clearly evident in the right panel.}
 \label{fig:MWmaps}
\end{figure}

In Section \ref{sec:metal_evo} we discuss the complexity of element injection and transport and how our simulations capture the key qualitative feature that metallicity statistics result from a competition between stochastic injection of SNe, which produce metal inhomogeneity, and mixing by turbulence, which smooths out the abundance variations. Notably we show that simulations with higher SN rate per unit volume produce higher abundance dispersion at the same mean metallicity. This relationship is even more clear for simulations in which the galaxy potential and gas surface density are fixed. For example at $\langle [\text{Fe}/\text{H}] \rangle = -2.5$ the [Fe/H] spread (measured as the $1\sigma$ of the distribution) is: 0.22, 0.27 and 0.55 for the low $\epsilon_*$, base model and high $\epsilon_*$, respectively. In Figure \ref{fig:MWmaps} we show a face-on view of the galaxy disc in the three models at the same $\langle [\mathrm{Fe}/\mathrm{H}] \rangle$. As expected, the model with the highest $\epsilon_*$ has a much more patchy distribution of metals at the same mean abundance. This is because at the same mean abundance (i.e., a similar amount of SNe) the individual SNR do not have as much time to spread and mix as they do when the rate is lower. This leads to a more patchy distribution of metals for the high $\epsilon_*$ model.

\begin{figure}
\centering
\includegraphics[width = 0.5\textwidth]{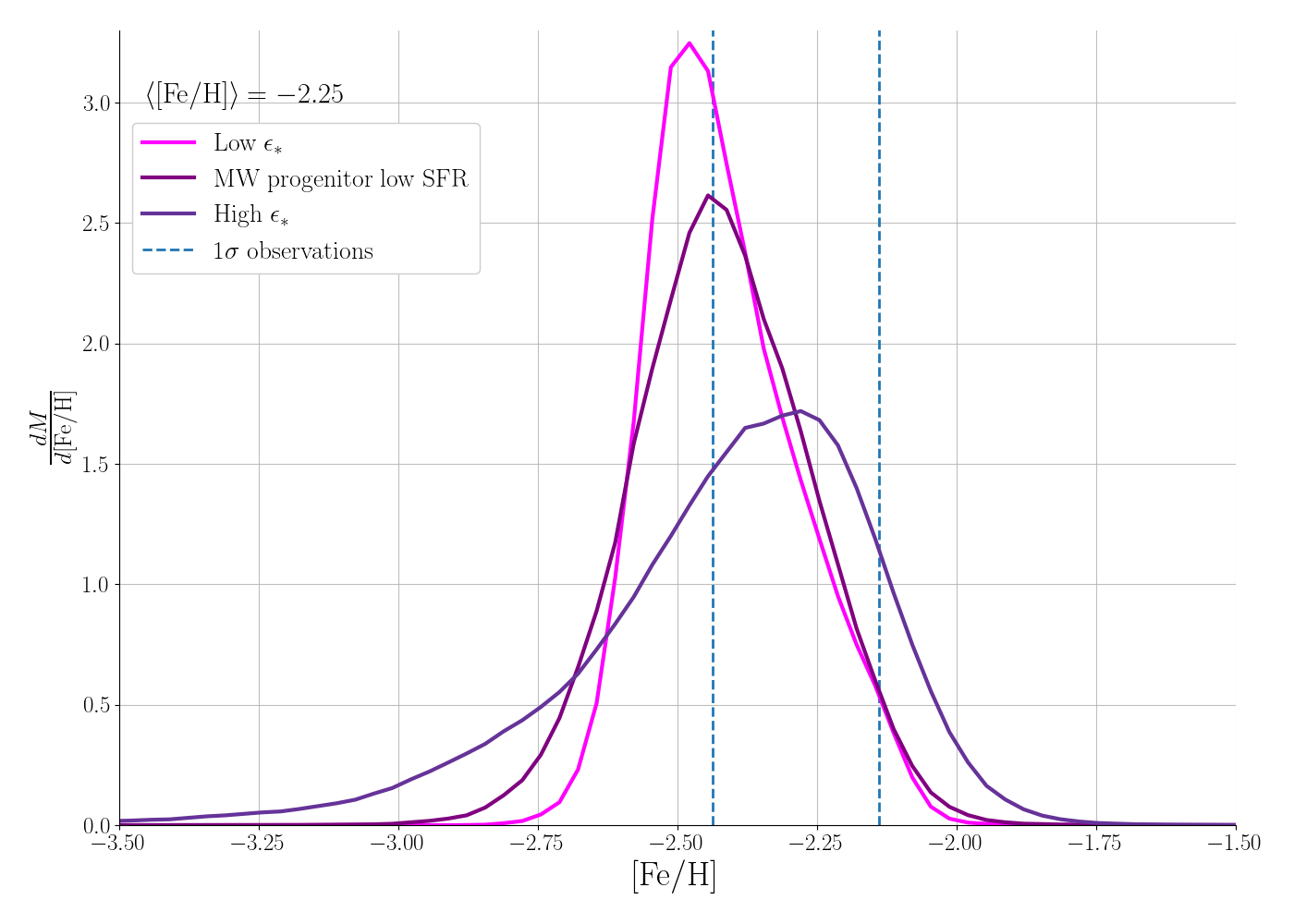}
 \caption{Comparison of the 1d distribution of [Fe/H] in the three models at mean [Fe/H] = -2.25. Also shown are the $1 \sigma$ spread of observations of [Fe/H]  at the same corresponding mean [Fe/H] metallicity (dashed lines). \label{fig:MW_spread}}
\end{figure}

To study the abundance variations with $\epsilon_*$ we consider the distribution of [Fe/H] in the cold, dense gas, which is selected in the same way as in Section \ref{fig:cold_dense_abundances}. In Figure \ref{fig:MW_spread} we compare the 1d distributions of [Fe/H] at $\langle [\mathrm{Fe}/\mathrm{H}] \rangle = -2.25$ in the three simulations. While the mean abundance of [Fe/H]  is similar in all three models, the shapes of the distributions are very different as well as their corresponding spreads. From Figure \ref{fig:MW_spread} we conclude that the high $\epsilon_*$ model is much less consistent with observations, thus reinforcing our conclusions about a preference for models with a modest SFR.

\FloatBarrier
\section{Results with intrinsic  SN yield variations}

Here we show the results of the three different models presented in Section~\ref{sec:stars}. These models are similar to those presented in
Section~\ref{sec:metal_evo}, although those models  assumed no intrinsic  SN yield variations. These chemically identical models are presented in Figure~\ref{fig:maps} and are used to study the mixing of Fe as a function of the global mean [Fe/H] abundance.  
However, comparison with observations, as shown in Figure~\ref{fig:cold_dense_abundances}, are aided by having two independent tracers that can isolate the rate of SN injection and the effects of turbulent mixing. We do this in this {\it Letter}  by introducing an intrinsic spread in yields from SN to SN, which is required by observations of metal poor stars as described in Appendix~\ref{app:obs}.

Figure~\ref{fig:MgFe_maps} shows  [Mg/Fe] slice maps of the for each of the three galaxy models
considered in this work. Individual SNR with varying yields can be easily spotted. The influence of the metal injection rate, which is driven by the SFR, can be clearly seen when comparing the size of individual remnants between different models. Remnants  are observed to reach larger physical sizes in MW progenitor models with  lower SFR  than remnants in the MW progenitor model with high SFR. This produces a higher degree of smoothing  in the spread of [Mg/Fe] of the gas at given $\langle [\text{Fe}/\text{H}] \rangle$ in the lower SFR models. One important feature of the comparison between our simulations and the observations in Figure~\ref{fig:cold_dense_abundances} is that we are directly matching how the spread in [Mg/Fe] evolves with $\langle [\text{Fe}/\text{H}] \rangle$, which is initially sensitive to the assumed SN-to-SN yield variance and is then subsequently smoothed out by turbulent diffusion. As we have argued throughout this {\it Letter}, the strength of turbulent diffusion depends on the SFR and, as such, can be used to derive constraints on the star formation conditions giving rise to metal poor stars.

\begin{figure}
    \centering
    \includegraphics[width = \textwidth]{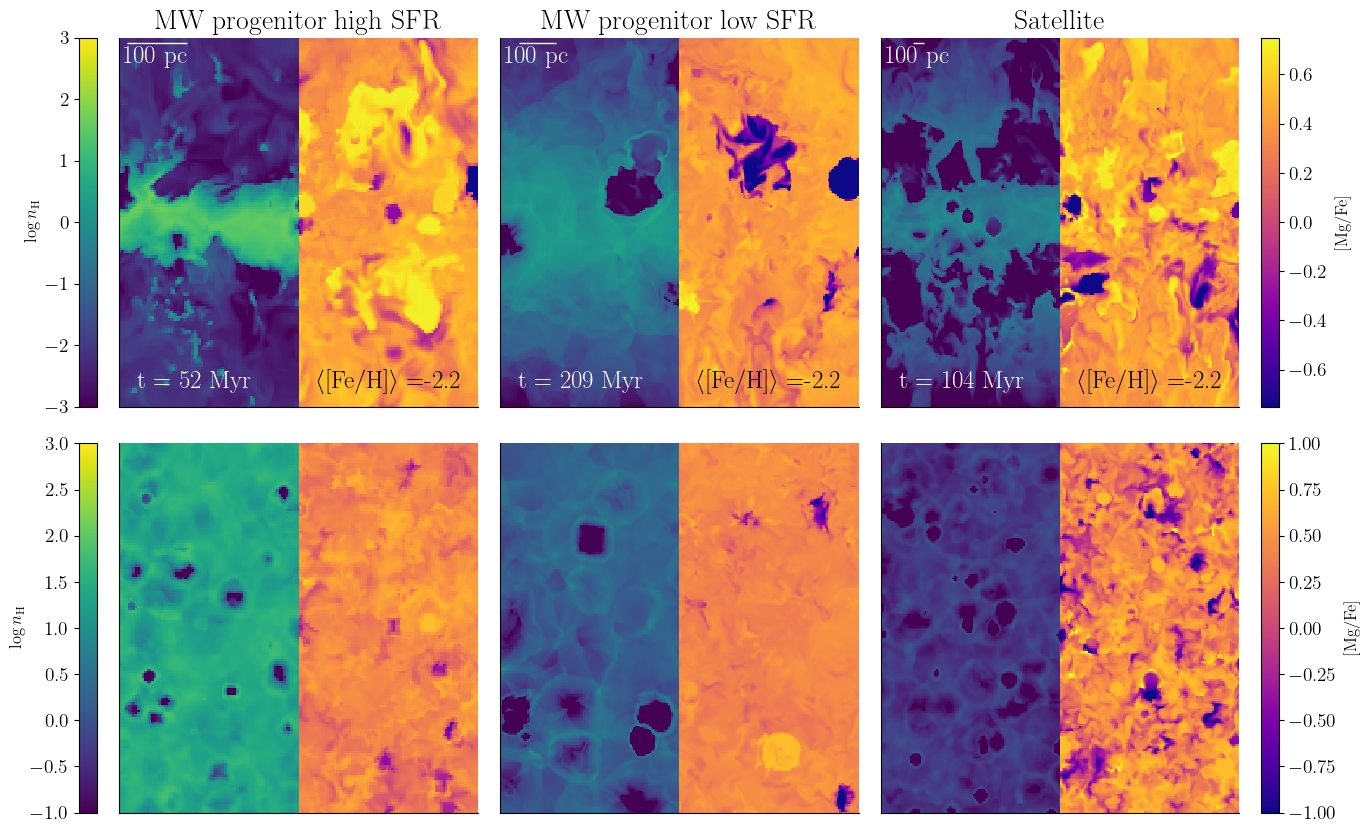}
    \caption{Same as Figure \ref{fig:maps} but showing the [Mg/Fe] abundance in place of the [Fe/H] abundance from the simulations with intrinsic SN yield variations described in Section~\ref{sec:metal_inj}  and empirically  derived in Appendix~\ref{app:obs}.}
    \label{fig:MgFe_maps}
\end{figure}

\end{document}